\documentclass[fullpage]{article}
\pdfoutput=1
\newcounter{mnotecount}[section]

\renewcommand{\themnotecount}{\thesection.\arabic{mnotecount}}

\newcommand{\mnote}[1]
{\protect{\stepcounter{mnotecount}}$^{\mbox{\footnotesize
$
\bullet$\themnotecount}}$ \marginpar{
\raggedright\tiny\em
$\!\!\!\!\!\!\,\bullet$\themnotecount: #1} }

\usepackage[utf8]{inputenc}
\usepackage{amsmath,amssymb}
\usepackage{amsmath}
\usepackage{bbm}
\usepackage{amsthm}
\usepackage{enumerate}
\usepackage{braket}
\usepackage{bm}
\usepackage[title]{appendix}
\usepackage{tikz}
\usepackage[none]{hyphenat}
\usepackage{geometry}
\usepackage{float}
\usepackage{fancyhdr}
\usepackage{lipsum}
\usepackage{braket}
\usepackage[labelfont=bf]{caption}
\theoremstyle{definition}
\usepackage{pifont}
\newcommand{\cmark}{\ding{51}}%
\newcommand{\xmark}{\ding{55}}%

\title{The Penrose Property with a Cosmological Constant}
\author{Peter Cameron\footnote{pjc96@cam.ac.uk}\;\\
Department of Applied Mathematics and Theoretical Physics\\ 
University of Cambridge\\ Wilberforce Road, Cambridge CB3 0WA, UK.}
\date{\today}
\begin{document}
\theoremstyle{definition}
\newtheorem{definition}{Definition}[section]
\newtheorem{example}[definition]{Example}
\newtheorem{prop}[definition]{Proposition}
\newtheorem{lemma}[definition]{Lemma}
\newtheorem{thm}[definition]{Theorem}
\newtheorem{cor}[definition]{Corollary}
\maketitle
\begin{abstract}
\noindent 
\noindent 
A spacetime satisfies the non-timelike boundary version of the Penrose property if the timelike future of any point on $\mathcal{I}^-$ contains the whole of $\mathcal{I}^+$. This property was first discussed for asymptotically flat spacetimes by Penrose, along with an equivalent definition (the finite version). In this paper we consider the Penrose property in greater generality. In particular we consider spacetimes with a non-zero cosmological constant and we note that the two versions of the property are no longer equivalent. In asymptotically AdS spacetimes it is necessary to re-state the property in a way which is more suited to spacetimes with a timelike boundary. We arrive at a property previously considered by Gao and Wald. Curiously, this property was shown to fail in spacetimes which focus null geodesics. This is in contrast to our findings in asymptotically flat and asymptotically de Sitter spacetimes. We then move on to consider some further example spacetimes (with zero cosmological constant) which highlight features of the Penrose property not previously considered. We discuss spacetimes which are the product of a Lorentzian and a compact Riemannian manifold. Perhaps surprisingly, we find that both versions of the Penrose property are satisfied in this product spacetime if and only if they are satisfied in the Lorentzian spacetime only. We also discuss the Ellis-Bronnikov wormhole (an example of a spacetime with more than one asymptotically flat end) and the Hayward metric (an example of a non-singular black hole spacetime).
\end{abstract} 

\section{Introduction}\label{Introduction}
This papers follows \cite{PCMD}, which discussed a property of spacetimes first introduced by Penrose in \cite{Penrose}. We will be interested in spacetimes which admit a conformal compactification. Following \cite{HawkingEllis}, we make the following definitions
\begin{definition}\label{defn:asymptotically simple}
A time- and space-orientable spacetime $(M,g)$ is 
\textit{asymptotically simple} if there is a strongly causal spacetime $(\tilde{M},\overline{g})$ and an embedding $\varphi:M\rightarrow\tilde{M}$ which embeds $M$ as a manifold with smooth boundary $\partial\overline{M}$ in $\tilde{M}$ such that:
\begin{enumerate}
\item there is a smooth function $\Omega$ on $\tilde{M}$, with $\Omega>0$ and $\varphi^*(\overline{g})=\Omega^2g$ on $M$; and
\item $\Omega=0$ and $d\Omega\neq0$ on $\partial\overline{M}$.
\end{enumerate}
\end{definition}
We shall write $\overline{M}\equiv M\cup \partial\overline{M}$ and refer to $(\overline{M},\overline{g})$ as the conformal compactification of $(M,g)$. Definition \ref{defn:asymptotically simple} ensures that such a conformal compactification exists. 

In an asymptotically simple spacetime, we define $\mathcal{I}^+$ to be elements of $\partial\overline{M}$ which are future endpoints of null geodesics and we define $\mathcal{I}^-$ to be elements of $\partial\overline{M}$ which are past endpoints of null geodesics. The union of these two sets will be referred to as the \textit{conformal boundary at infinity}, denoted $\mathcal{I}$. Note that $\mathcal{I}^+$ and $\mathcal{I}^-$ may have non-empty intersection. For example in the compactification of anti-de Sitter spacetime (Section \ref{The Penrose Property in Asymptotically Anti-de Sitter Spacetime}), we have $\mathcal{I}=\mathcal{I}^+=\mathcal{I}^-$.

In \cite{PCMD} the spacetimes considered satisfied an additional condition (referred to there as \textit{asymptotically empty and simple}). 
\begin{definition}
A time- and space-orientable spacetime $(M,g)$ is 
\textit{asymptotically flat} if it is asymptotically simple and the Ricci tensor of $g$ vanishes on some neighbourhood of $\mathcal{I}$ in $\overline{M}$.
\end{definition}
In this paper we will also consider spacetimes which approach de Sitter or anti-de Sitter at infinity. To do this, we make the following definitions:

\begin{definition}\label{defn:asymptotically de Sitter}
A time- and space-orientable spacetime $(M,g)$ is 
\textit{asymptotically de Sitter} if it is asymptotically simple and $R_{ab}=\Lambda g_{ab}$ on some neighbourhood of $\mathcal{I}$ in $\overline{M}$, for some $\Lambda>0$.
\end{definition}
\begin{definition}\label{defn:asymptotically anti-de Sitter}
A time- and space-orientable spacetime $(M,g)$ is 
\textit{asymptotically anti-de Sitter} if it is asymptotically simple and $R_{ab}=\Lambda g_{ab}$ on some neighbourhood of $\mathcal{I}$ in $\overline{M}$, for some $\Lambda<0$.
\end{definition}

In \cite{PCMD}, the focus was on the following property:

\begin{definition}[Penrose property - non-timelike boundary version]\label{defn:Penroseproperty}
An asymptotically flat or asymptotically de Sitter spacetime, $(M,g)$, with conformal compactification $(\overline{M},\overline{g})$ satisfies the \textit{non-timelike boundary version of the Penrose property} if any $p\in\mathcal{I}^-$ and any $q\in\mathcal{I}^+$ can be connected by a smooth timelike curve.
\end{definition}
This definition will be adapted in the case of spacetimes with timelike boundary (see Definition \ref{defn:PenrosePropertyAdS}). 

Besides defining this property, Penrose also shows in \cite{Penrose} that there is an equivalent version which does not make reference to a conformal compactification but instead relates to endless timelike curves. Before continuing, it will be helpful to define the following sets (for some $p\in\overline{M}$):
\begin{equation}
    \begin{split}
        J^+(p)&=\{q\in\overline{M}|\exists\text{ a smooth future-directed causal curve in $\overline{M}$ from $p$ to $q$}\}\\
        J^-(p)&=\{q\in\overline{M}|p\in J^+(q)\}\\
        I^+(p)&=\{q\in\overline{M}|\exists\text{ a smooth future-directed timelike curve in $\overline{M}$ from $p$ to $q$}\}\\
        I^-(p)&=\{q\in\overline{M}|p\in I^+(q)\}.
    \end{split}
\end{equation}

In these definitions we include curves in the boundary of $\overline{M}$ and determine the causal nature of such curves using the conformal metric, $\overline{g}$, which extends to $\partial\overline{M}$. For non-compact spacetimes, $(M,g)$, we can define the same sets as subsets of $M$ rather than as subsets of $\overline{M}$.

As in \cite{Penrose} we will concern ourselves only with inextendible timelike curves in $M$ whose images in the conformal compactification $\overline{M}$ are contained in the domain of outer dependence $\mathcal{D}=J^-(\mathcal{I}^+)\cap J^+(\mathcal{I}^-)$.


\begin{definition}[Penrose Property - finite version]\label{defn:Penrosefinite} An asymptotically simple spacetime $(M,g)$ satisfies the \textit{finite version of the Penrose property} if for any endless timelike curves $\lambda,\nu\subset J^-(\mathcal{I}^+)\cap J^+(\mathcal{I}^-)$, there exists $p\in\lambda$ and $q\in\nu$ such that $p$ can be connected to $q$ by a future pointing timelike curve in $M$. 
\end{definition}
The condition that $\lambda,\nu\subset J^-(\mathcal{I}^+)\cap J^+(\mathcal{I}^-)$ ensures that these curves do not cross any event horizons.

\begin{thm}[Penrose: Theorem IV.4 \cite{Penrose}]\label{thm:penroseequivalence}
The two versions of the Penrose property, Definitions \ref{defn:Penroseproperty} and \ref{defn:Penrosefinite}, are equivalent for asymptotically flat spacetimes. 
\end{thm}

Penrose's motivation for studying the Penrose property was the following corollary. We use the notation $g_1\leq g_2$ for two metrics defined on the same manifold to mean that the lightcones of $g_1$ do not extend outside the lightcones of $g_2$. In other words $g_1\leq g_2$ is the condition that
\begin{equation}\label{eqn:metricineq}
    g_1(T,T)>0\implies g_2(T,T)>0
\end{equation}
at every point and for every tangent vector $T$.\footnote{For consistency, we follow \cite{PCMD} and \cite{Penrose} in using the `mostly minus' signature $(+,-,...,-)$. We also set $c=G=1$ throughout.}\footnote{This is slightly different to \cite{Penrose} which denotes this condition by $g_1<g_2$. Our definition seems more appropriate since equation (\ref{eqn:metricineq}) is clearly satisfied if $g_1=g_2$.}
\begin{cor}[Penrose: Theorem IV.5 \cite{Penrose}]\label{cor:backgroundMink}
Let $(\overline{M},\overline{g})$ be the conformal compactification of an asymptotically flat spacetime which satisfies the non-timelike boundary version of the Penrose property and let $\mathcal{I}$ denote its conformal boundary at infinity. Suppose a subset of Minkowski spacetime that includes a neighbourhood of $i^0$ can be conformally embedded into a subset of $\overline{M}$, as in Definition \ref{defn:asymptotically simple}, which contains $i^0$ and has $\mathcal{I}$ as its conformal boundary at infinity. Then there is no neighbourhood of $i^0$ on which the inequality $\overline{g}\leq\overline{\eta}$ holds.

%
\end{cor}
\textbf{Proof:} Suppose there does exist a neighbourhood, $U$, of $i^0$, on which the condition $\overline{g}\leq\overline{\eta}$ holds. Minkowski spacetime does not satisfy the non-timelike boundary version of the Penrose property \cite[Proposition 3.3]{PCMD}, hence we can choose $p\in\mathcal{I}^-$ and $q\in\mathcal{I}^+$ to be points which cannot be connected by a $\overline{\eta}$-timelike curve and which are sufficiently close to $i^0$ so that any $\overline{g}$ causal curve between them must lie entirely in $U$. Since $(\overline{M},\overline{g})$ satisfies the non-timelike version of the Penrose property, we therefore have a $\overline{g}$-timelike curve $\lambda\subset U$ from $p$ to $q$. The condition $\overline{g}\leq\overline{\eta}$ ensures that this curve is also timelike with respect to the metric $\overline{\eta}$. This is a contradiction. \qedsymbol

The inequality $\overline{g}\leq\overline{\eta}$ arises as a consistency condition for theories of quantum gravity defined with respect to a fixed ``background Minkowski spacetime'' \cite{Penrose}. If one tries to write the ``phyiscal'' field operators in terms of fields defined on the background Minkowski spacetime (and hence propagating inside the background Minkowski lightcones), then this condition ensures that the physical fields cannot propagate at super-luminal speeds with respect to the metric $g$.

The main results of $\cite{PCMD}$ and $\cite{Penrose}$ were to show that the two (equivalent) versions of the Penrose property appear to be satisfied only in positive mass spacetimes in low dimensions. In particular, the follow theorem relating to the Schwarzschild spacetime was proved:

\begin{definition}\label{defn:Schwarzschildhigherdim} The Schwarzschild metric in $d+1$ dimensions ($d\geq3$) is given by \cite{EmparanReall}
\begin{equation}\label{eqn:higherdimSchwarzschild}
    ds^2=V(r)dt^2-\frac{dr^2}{V(r)}-r^2d\omega_{d-1}^2
\end{equation}
where $V(r)=1-\frac{\mu}{r^{d-2}}$ and $d\omega_{d-1}^2$ denotes the round metric on a unit $(d-1)$--dimensional sphere $S^{d-1}$. We have also introduced the mass parameter, $\mu$, which is related to the ADM mass \cite{ADM}, $m$, by
\begin{equation}
    \mu=\frac{16\pi m}{(d-1)A_{S^{d-1}}}
\end{equation}
where $A_{S^{d-1}}$ denotes the area of $S^{d-1}$.

If $m<0$ then we consider this metric on the region $r>0$. If $m>0$ then we consider only the \textit{exterior region} $r^{d-2}>\mu$. In this case, a change of co-ordinates can be made which allows the metric to be extended across the surface $r^{d-2}=\mu$ (see \cite{HawkingEllis} which does this in 3+1 dimensions, with the higher dimensional case treated similarly), however we will not consider this here. \end{definition}

\textbf{Theorem A \cite{PCMD}:}
The Penrose property is satisfied by Schwarzschild spacetime of mass $m$ and varying spacetime dimension according to the table below
\begin{center}
    \begin{tabular}{|c|c|c|}\hline
  Spacetime dimension & $m>0$ & $m\leq0$\\
   \hline
 $3$ & \cmark & \xmark\\ 
 $4$ & \cmark & \xmark\\  
 $\geq5$ & \xmark &  \xmark\\
 \hline
    \end{tabular}
    \label{table:penroseproperty}
\end{center}

\begin{figure}
    \centering
    \includegraphics[scale=0.2]{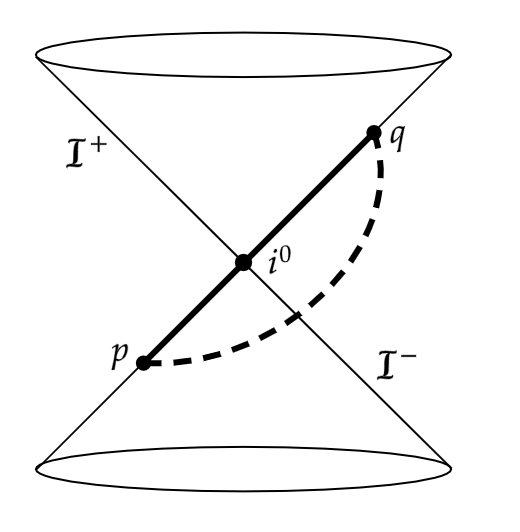}
    \caption{The non-timelike boundary version of the Penrose property asks whether the properties of the spacetime near $i^0$ allow all points on $\mathcal{I}^-$ and $\mathcal{I}^+$ to be connected using timelike curves. This figure shows $\mathcal{I}^-$ and $\mathcal{I}^+$ as the past and future lightcones emanating from $i^0$, as described in \cite{AshtekarHansen}.}
    \label{fig:spatialinfinity}
\end{figure}

This perhaps surprising result can be understood intuitively as follows. The Minkowski metric in $(d+1)$--dimensions is defined on $\mathbbm{R}^{1,1}\times S^{d-1}$. This can be compactified as in \cite[Section 3]{PCMD} to give a spacetime on $V\times S^{d-1}$, where $V\subset\mathbbm{R}^{1,1}$ is a manifold with boundary. In this compactification, certain pairs of points on $\mathcal{I}^-$ and $\mathcal{I}^+$ which project to antipodal points on $S^{d-1}$ cannot be connected by timelike curves \cite[Proposition 3.3]{PCMD}. They can however be connected by a null curve passing through $i^0$, as shown in Figure \ref{fig:spatialinfinity}. As explained in \cite{AshtekarHansen}, spacetimes with positive ADM mass can be thought of as containing a point mass at $i^0$. If the ADM mass is positive, we expect this point mass to focus null geodesics. This may allow all antipodal points on $\mathcal{I}^-$ and $\mathcal{I}^+$ to be timelike connected. However, this focusing is offset by the time delay of null geodesics in positive mass spacetimes which may prevent us from timelike connecting points near spatial infinity. Whether or not the non-timelike boundary version of the Penrose property is satisfied depends on the interplay between these two effects.

The exterior region of the $(d+1)$--dimensional positive mass Schwarzschild spacetime (Definition \ref{defn:Schwarzschildhigherdim}) is defined on the submanifold of $\mathbbm{R^{d,1}}$ given by the restriction $\{r^{d-2}>\mu\}$. We can also define a flat Minkowski metric, $ds^2_{Mink}$, on the same manifold by setting $\mu=0$ in equation (\ref{eqn:higherdimSchwarzschild}). These line elements then satisfy
\begin{equation}
    ds^2\leq ds^2_{Mink}.
\end{equation}

In higher dimensions, we can compactify these metrics using the same embedding map, so we obtain
\begin{equation}\label{eqn:compactcomparison}
    \overline{ds}^2\leq \frac{\Omega^2}{\Omega^2_{Mink}}\overline{ds}^2_{Mink}
\end{equation}
where $\Omega^2$ and $\Omega^2_{Mink}$ are the compactification functions associated with $\overline{ds}^2$ and $\overline{ds}^2_{Mink}$ respectively (see Definition \ref{defn:asymptotically simple}).

This tells us that in higher dimensions, positive mass Schwarzschild null geodesics are delayed relative to null geodesics corresponding to the metric obtained by setting $\mu=0$. Along with the known failure of the non-timelike boundary version of the Penrose property in Minkowski spacetime, this allows us to deduce that this property also fails in higher dimensional positive mass Schwarzschild. 3+1 dimensions is a special case since any Schwarzschild null geodesic reaching infinity is infinitely delayed relative to Minkowski null geodesics. This is a result of the logarithmic divergence of the tortoise co-ordinate $r_*$ away from the radial co-ordinate $r$ and is unique to 4 dimensions. As a result, different compactified co-ordinates are required which means the comparison in equation (\ref{eqn:compactcomparison}) is no longer valid. In fact, it is the relative time advancement of null geodesics near infinity when compared to those passing nearer to the singularity which is now important. This relative time advancement becomes infinite as the impact parameter $R\rightarrow\infty$. This allows us to find null geodesics with endpoints arbitrarily near to spatial infinity. This feature is used to show that the non-timelike boundary version of the Penrose property is satisfied \cite[Theorem 4.1]{PCMD}\cite{Penrose}. 

The time advancement of null geodesics in negative mass higher dimensional Schwarzschild spacetime reverses inequality (\ref{eqn:compactcomparison}). This would appear to help us to timelike connect points near spatial infinity. However, this effect turns out to be insufficient to overcome the de-focusing caused by the negative point mass at $i^0$, and the non-timelike boundary version of the Penrose property is not satisfied. This is discussed further in \cite{PCMD}. The corresponding relative delay of geodesics with larger impact parameter is the important feature in the higher dimensional proof of the positive mass theorem in \cite{PC}. The 3+1 dimensional positivity of mass proof of Penrose, Sorkin and Woolgar \cite{PSW} relies on the light ray divergence described above and the failure of the non-timelike boundary version of the Penrose property in 3+1 dimensional negative mass spacetimes. 

In this paper we consider spacetimes with non-zero cosmological constant. We discuss the two versions of the Penrose property (or appropriate modifications) in simple examples and consider whether or not they remain equivalent.

We begin by considering spacetimes which are asymptotically de Sitter. In Section \ref{De Sitter Spacetime} we show that the non-timelike boundary version of the Penrose property is not satisfied in de Sitter spacetime of any dimension. However, in Section \ref{Schwarzschild-de Sitter} we show that the Schwarzschild-de Sitter black hole satisfies the non-timelike boundary version of the Penrose property in any spacetime dimension $\geq4$ if and only if the mass is strictly positive. 

In Section \ref{The Penrose Property in Asymptotically Anti-de Sitter Spacetime} we consider spacetimes which are asymptotically anti-de Sitter. We consider causality in pure AdS spacetime and provide justification for an appropriate analogue of the non-timelike boundary version of the Penrose property for spacetimes with a timelike boundary. We recall the results of the Gao-Wald theorem \cite[Theorem 2]{GW} which states that our new timelike boundary version of the Penrose property fails for spacetimes which focus null geodesics. Furthermore, we show that our property is satisfied by the AdS-Schwarzschild spacetime if and only if the mass parameter is strictly negative. These results are in contrast with those obtained in asymptotically flat and asymptotically de Sitter spacetimes, where the non-timelike boundary version of the Penrose property appears to be associated with 3 or 4 dimensional positive mass spacetimes which focus null geodesics. 

The following theorem summarises the relationship between the finite version of the Penrose property (Definition \ref{defn:Penrosefinite}) and the appropriate boundary version (Definition \ref{defn:Penroseproperty} for spacetimes with $\Lambda\geq0$ or Definition \ref{defn:PenrosePropertyAdS} if $\Lambda<0$):
\begin{thm}\label{thm:equivalence} In the presence of a cosmological constant, the finite version of the Penrose property is related to the appropriate boundary version according to the following table.
\begin{center}
    \begin{tabular}{|c|c|c|c|}\hline
    & $\Lambda=0$ & $\Lambda>0$ & $\Lambda<0$\\
     \hline
     Finite Version $\iff$ Appropriate Boundary Version? & \cmark & \cmark & \xmark \\
     \hline
    \end{tabular}
    \label{table:penroseproperties}
\end{center}
\end{thm}

We also summarise the appropriate boundary version of the Penrose property in various cosmological spacetimes in the following theorem.\\
\\
\textbf{Theorem B:} The appropriate boundary version of the Penrose property is satisfied according to the following table.
\begin{center}
    \begin{tabular}{|c|c|c|c|c|c|c|c|c|c|}\hline
    & \multicolumn{3}{c|}{$\Lambda=0$} & \multicolumn{3}{c|}{$\Lambda>0$} & \multicolumn{3}{c|}{$\Lambda<0$}\\
    \hline
  Spacetime dimension & $m>0$ & $m=0$ &$m<0$ & $m>0$ & $m=0$ & $m<0$ & $m>0$ & $m=0$ & $m<0$\\
  \hline
 $3$ & \cmark & \xmark &\xmark & - &\xmark &- &- &\xmark &- \\ 
 $4$ & \cmark & \xmark &\xmark & \cmark & \xmark & \xmark & \xmark & \xmark & \cmark \\  
 $\geq5$ & \xmark & \xmark &\xmark & \cmark & \xmark & \xmark & \xmark & \xmark & \cmark \\
 \hline
    \end{tabular}
\label{table:thmB}
\end{center}
where a dash indicates that this spacetime has not been considered here.

Finally, we consider some example spacetimes which highlight features of the Penrose property which were not discussed in \cite{PCMD}. We begin in Section \ref{Product Spacetimes} by discussing the Penrose property in spacetimes of the form $(M',g')\times(M'',g'')$, where $(M',g')$ is a Lorentzian manifold and $(M'',g'')$ is a compact Riemannian manifold. Intuitively, one might expect both versions of the Penrose property to be more restrictive on this product spacetime than on the spacetime $(M',g')$ alone. We find however that they are in fact equivalent. In Section \ref{Ellis-Bronnikov Wormhole} we consider the Ellis-Bronnikov wormhole spacetime. This is a spacetime with two asymptotically flat ends connected by a traversable wormhole. We find that the results for points on $\mathcal{I}^\pm$ in the same universe are the same as for Minkowski spacetime, however for points located in different universes the result is a more complicated function of the angular separation and time of flight (Definition \ref{defn:timeofflight}) along a curve between the two points. In Section \ref{Hayward Metric} we consider the Hayward metric in 3+1 dimensions. This metric describes a non-singular black hole spacetime with the same asymptotic behaviour as Schwarzschild. Consequently we find that this spacetime also satisfies the Penrose property, highlighting the fact that this is a property of neighbourhoods of $i^0$ and does not require the existence of a curvature singularity. 

\section{Spacetimes with Cosmological Constant}\label{Spacetimes with Cosmological Constant}

In \cite{Penrose}, Penrose introduces two properties of spacetime (Definitions \ref{defn:Penroseproperty} and \ref{defn:Penrosefinite}) which he shows to be equivalent (Theorem \ref{thm:penroseequivalence}). In this section we consider these properties and their equivalence in the presence of a cosmological constant. 

\subsection{De Sitter Spacetime}\label{De Sitter Spacetime}
De Sitter spacetime in $d+1$ dimensions is the isometrically embedded submanifold of Minkowski spacetime of signature $(d+1,1)$, which we denote $\text{Mink}_{d+1,1}$, defined by
\begin{equation}\label{eqn:dSsubspace}
    -x_0^2+\sum_{i=1}^{d+1}x_i^2=l^2
\end{equation}
where $(x_0,x_1,...,x_{d+1})$ are a standard Cartesian co-ordinate system on $\mathbbm{R}^{d+1,1}$ and the radius of curvature, $l$, is a positive constant which is related to the cosmological constant by
\begin{equation}
    \Lambda=\frac{d(d-1)}{2l^2}
\end{equation}

De Sitter spacetime is topologically $\mathbbm{R}\times S^d$. The metric expressed in static co-ordinates is 
\begin{equation}\label{eqn:AdS}
    \begin{split}
        ds^2=\left(1-\frac{r^2}{l^2}\right)dt^2-\left(1-\frac{r^2}{l^2}\right)^{-1}dr^2-r^2d\omega^2_{d-1}
    \end{split}
\end{equation}

\begin{figure} 
    \centering
    \includegraphics[scale=0.3]{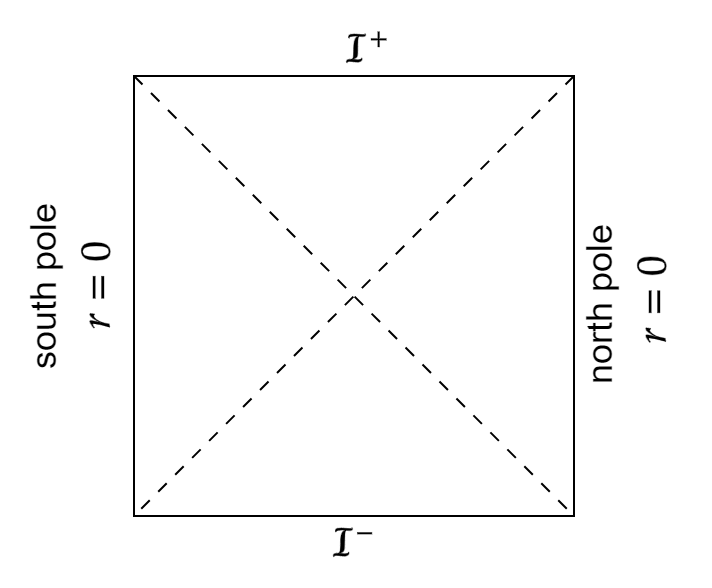}
    \caption{Penrose diagram of de Sitter spacetime. This spacetime is topologically $\mathbbm{R}\times S^d$, so each point on this diagram (with the exception of the north and south poles) represents a manifold which is topologically $S^{d-1}$. The dotted lines represent the cosmological horizon.}
    \label{fig:deSitter}
\end{figure}

The Penrose diagram of this spacetime is shown in Figure \ref{fig:deSitter}. Note that the hypersurfaces $\mathcal{I}^\pm$ are spacelike. 

\begin{thm}\label{thm:dSPenroseproperty}
Compactfied de Sitter spacetime does not satisfy the non-timelike boundary version of the Penrose property. Moreover, for any $p\in\mathcal{I}^-$, the only point on $\mathcal{I}^+$ which cannot be reached from $p$ by a smooth timelike curve is the antipodal point.
\end{thm}

\textbf{Proof}: If we choose co-ordinates such that $p\in\mathcal{I}^-$ lies at the south pole then it is clear from Figure \ref{fig:deSitter} that this point is not timelike connected to the north pole on $\mathcal{I}^+$. 

\begin{figure} 
    \centering
    \includegraphics[scale=0.3]{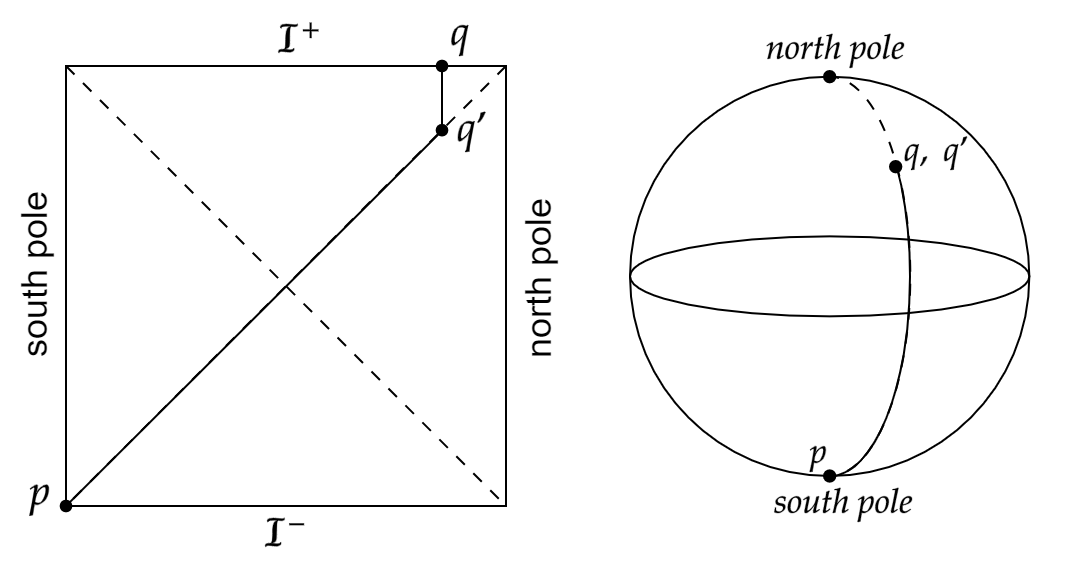}
    \caption{We choose a null geodesic from the south to north pole which passes through the point, $q'$, which has the same co-ordinates on $S^d$ as $q$ (although they have different $t$ co-ordinates). This null geodesic is a great circle when projected onto $S^d$. We then follow a timelike path from $q$ to $q'$ which remains at a constant position on $S^d$.}
    \label{fig:deSittercompositepath}
\end{figure}

Now suppose $q\in\mathcal{I}^+$ does not lie at the north pole. The following construction is illustrated in Figure \ref{fig:deSittercompositepath}. Consider the null geodesic with past endpoint at $p$ and whose angular co-ordinates pass through those of $q$. In other words, in the $S^d$ part of the compactified spacetime we follow the great circle from the south pole to the point which is the projection of $q$ onto $S^{d}$. We denote by $q'$ the point we reach in the full spacetime (so the projection at $q'$ onto $S^d$ is the same as the projection from $q$). At $q'$ we switch to a timelike path ending at $q$ which remains at a constant point on $S^d$. We can then modify the full path so as to obtain a smooth timelike curve from $p$ to $q$ \cite{DiffTop}. \qedsymbol

This behaviour is similar to that observed for Minkowski spacetime \cite[Proposition 3.3]{PCMD} where, as described in Section \ref{Introduction}, the only points on $\mathcal{I}^-$ and $\mathcal{I}^+$ which cannot be timelike connected are certain pairs which project to antipodal points on the $S^{d-1}$ part of the compactified spacetime $V\times S^{d-1}$, where $V\subset\mathbbm{R}^{1,1}$ is compact. 

It is also trivial to see that the finite version of the Penrose property is not satisfied in de Sitter spacetime.

\begin{thm}
The finite version of the Penrose property is not satisfied in de Sitter spacetime in $(d+1)$ dimensions for any $d\geq1$.
\end{thm}
\textbf{Proof:} Consider two timelike curves, one remaining at the north pole and the other at the south pole. It is clear from Figure \ref{fig:deSitter} that these curves cannot be timelike connected. 

In fact, we can also see the failure of the two versions of the Penrose property in de Sitter spacetime by recalling its definition as an embedded submanifold in Minkowski spacetime of one dimension higher. This argument is based on the one given in \cite{Penrose} to show that Minkowski spacetime does not satisfy either version of the Penrose property. 

Consider the hyperbola in $\text{Mink}_{d+1,1}$ defined by \begin{equation}\label{eqn:hyperboladesitter}
    \begin{split}
        x_2=x_3=...=x_{d+1}&=0\\
        -x_0^2+x_1^2&=l^2
    \end{split}
\end{equation}

Note that this hyperbola satisfies equation (\ref{eqn:dSsubspace}) and hence its branches define endless timelike curves in de Sitter spacetime. These curves are everywhere spacelike separated, so we see that the finite version of the Penrose property fails. Moreover, they become null as we approach their endpoints. It follows that these endpoints lie on $\mathcal{I}^\pm$ in compactified de Sitter spacetime and we conclude that the past endpoint of one branch cannot be timelike connected to the future endpoint of the other. 

More generally, we see that the finite and non-timelike boundary versions of the Penrose property are equivalent in asymptotically de Sitter spacetime. 

\begin{thm}\label{thm:dsequivalence}
For asymptotically de Sitter spacetimes, the finite and non-timelike boundary versions of the Penrose property are equivalent.
\end{thm}
\textbf{Proof:} The proof of this is identical to the proof of Theorem IV.4 in \cite{Penrose} and we refer the reader to this reference. \qedsymbol 

Theorem also \ref{thm:dSPenroseproperty} leads to the following analogue of Theorem IV.5 in \cite{Penrose}:

\begin{thm}\label{thm:backgrounddS}
Let $(\overline{M},\overline{g})$ be the conformal compactification of an asymptotically de Sitter spacetime which satisfies the non-timelike boundary version of the Penrose property. Then there is no compactification of de Sitter spacetime which has $\mathcal{I}=\mathcal{I}^-\cup\mathcal{I}^+$ as its conformal boundary and $\overline{g}\leq\overline{g}_{dS}$ on $J^-(\mathcal{I}^+)\cap J^-(\mathcal{I}^-)$.
\end{thm}
\textbf{Proof:} Suppose such a compactification of de Sitter spacetime exists and let $\lambda\in J^-(\mathcal{I}^+)\cap J^-(\mathcal{I}^-)$ be a timelike curve between antipodal points on $\mathcal{I}^-$ and $\mathcal{I}^+$. Then the inequality $\overline{g}\leq\overline{g}_{dS}$ implies that $\lambda$ is timelike with respect to the compactified de Sitter metric. This contradicts Theorem \ref{thm:dSPenroseproperty}. \qedsymbol

\subsection{Schwarzschild-de Sitter}\label{Schwarzschild-de Sitter}
From Theorem \ref{thm:backgrounddS}, we see that in order to rule out theories of quantum gravity based on a fixed background de Sitter spacetime (as is done in \cite{Penrose} for theories defined with respect to a fixed background Minkowski spacetime) it is sufficient to find a physically relevant example which satisfies the non-timelike boundary version of the Penrose property. Inspired by \cite{Penrose}, we consider the Schwarzschild-de Sitter spacetime.

The Schwarzschild-de Sitter metric in $d+1$ dimensions is given by \cite{SdS}
\begin{equation}\label{eqn:SdS}
    \begin{split}
         ds^2&=V(r)dt^2-\frac{dr^2}{V(r)}-r^2d\omega_{d-1}^2\\
V(r)&=1-\frac{\mu}{r^{d-2}}-\frac{r^2}{l^2}
\end{split}
\end{equation}
We have also introduced the \textit{mass parameter}, $\mu$. We call $\mu$ the mass parameter because it is related by
\begin{equation}
    \mu=\frac{16\pi m}{(d-1)A_{S^{d-1}}}
\end{equation}
to the leading order term, $m$, in the Abbott-Deser mass \cite{AD} calculated just inside the cosmological horizon for a black hole with event horizon at $r_b<<l$.

\begin{figure} 
    \centering
    \includegraphics[scale=0.3]{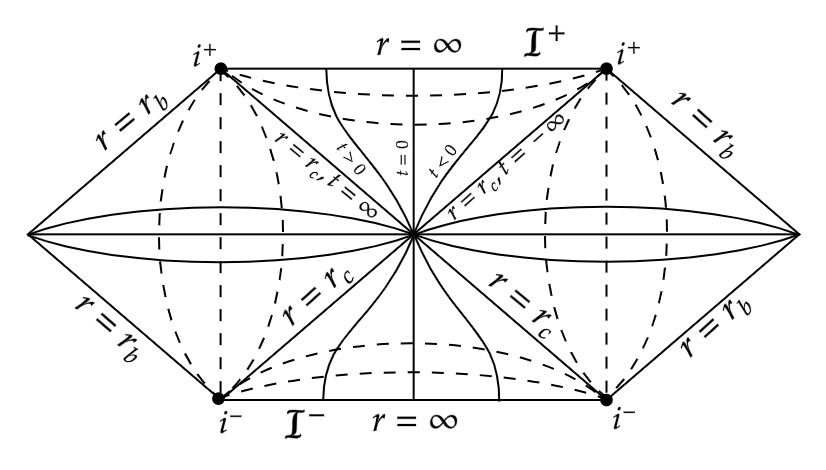}
    \caption{Portion of the Penrose diagram for the exterior ($r>r_b$) region of the Schwarzschild-de Sitter spacetime in the case where there are cosmological and event horizons which do not coincide. For $r_b<r<r_c$, constant $t$ surfaces are spacelike while constant $r$ surfaces are timelike. When $r>r_c$, constant $r$ surfaces are spacelike and constant $t$ surfaces are spacelike. See \cite{SchwarzschilddeSitter} for more details.}
    \label{fig:SchwarzschilddeSitter}
\end{figure}

\begin{figure}
    \centering
    \includegraphics[scale=0.3]{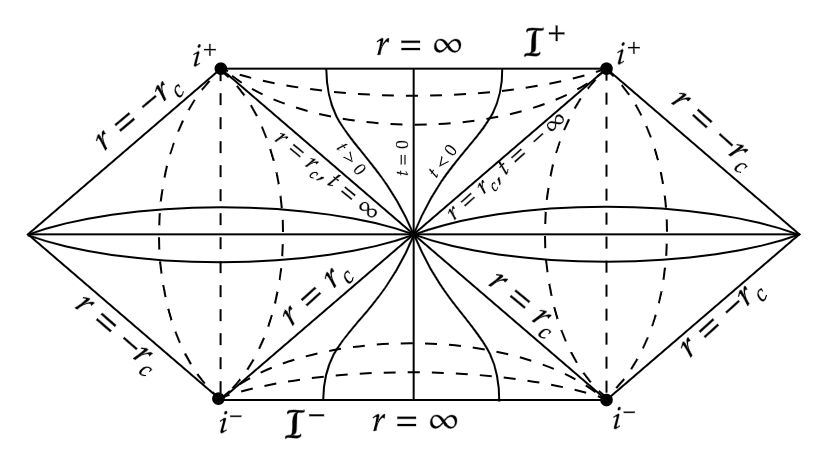}
    \caption{Portion of the Penrose diagram for the Nariai spacetime. This spacetime is the extremal case of Schwarzschild-de Sitter where $r_c=r_b$. It is non-singular at $r=0$ and can be extended to $r<0$.}
    \label{fig:nariai}
\end{figure}

We begin by considering the positive mass case $\mu>0$ and assume that $V(r)$ has two positive roots at $r_c\geq r_b>0$. In this case there is a cosmological horizon at $r=r_c$ and a black hole horizon at $r=r_b$. The Penrose diagram for the region $r>r_b$ is as shown in Figure \ref{fig:SchwarzschilddeSitter} (assuming $r_c> r_b>0$). The extremal case where $V(r)$ has a double root is known as the Nariai solution \cite{Nariai1950}\cite{Nariai1951}. In this case the spacetime is non-singular and can be extended through $r=0$. A portion of the Penrose diagram for this extension is shown in Figure \ref{fig:nariai}. 

\begin{thm}\label{thm:SchwarzschilddeSitter}
The non-timelike boundary version of the Penrose property is satisfied by the Schwarzschild-de Sitter spacetime with positive mass parameter $\mu>0$ in $d+1$ dimensions for all $d\geq3$.
\end{thm}

\textbf{Proof}: For $r>r_c$ we have $V(r)<0$. Suppose we follow a null curve of constant $t$ in the equatorial plane $\theta_1=...=\theta_{d-2}=\pi/2$. Let $\Delta\phi$ denote the change in the co-latitude $\phi$ along the section of this curve from $r=r_c$ to $r=\infty$. We have
\begin{equation}\label{eqn:deltaphi}
\begin{split}
    -\frac{dr^2}{V(r)}&=r^2d\phi^2\\
    \implies\Delta\phi&=\int_{r_c}^\infty\frac{dr}{r\sqrt{\frac{r^2}{l^2}+\frac{\mu}{r^{d-2}}-1}}\\
    &=\int_1^\infty\frac{dy}{y\sqrt{\frac{r_c^2}{l^2}\left(y^2-\frac{1}{y^{d-2}}\right)+\frac{1}{y^{d-2}}-1}}\\
    \end{split}
\end{equation}
where we have made the substitution $y=r/r_c$ and used the fact that $r_c$ satisfies $V(r_c)=0$.

\begin{figure} 
    \centering
    \includegraphics[scale=0.3]{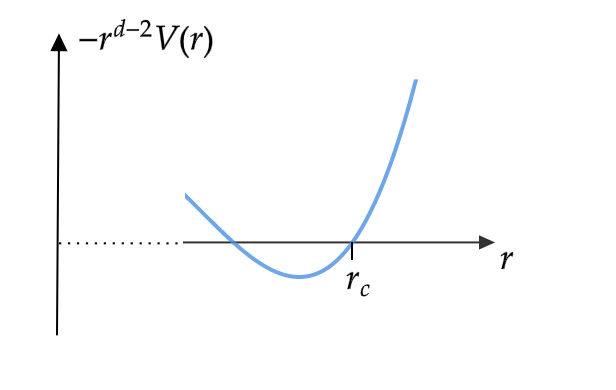}
    \caption{A portion of the function $-r^{d-2}V(r)=\mu-r^{d-2}+r^d/l^2$ is plotted (assuming $\mu>0$). If $\mu$ is increased slightly then $r_c$, the value of $r$ at the largest root of $r^{d-2}V(r)$, will decrease.}
    \label{fig:r^d-2V(r)}
\end{figure}

It will be convenient to consider $\Delta\phi$ as a function of $\mu$. An illustration of the graph of 
\begin{equation}
-r^{d-2}V(r)=\mu-r^{d-2}+r^d/l^2
\end{equation}
is shown in Figure \ref{fig:r^d-2V(r)}. Increasing $\mu$ infinitesimally will shift this graph upwards, which in turn will decrease $r_c$ (the largest root of this graph). This tells us that
\begin{equation}
    \frac{\partial r_c^2}{\partial \mu}<0
\end{equation}

We also see that
\begin{equation}\label{eqn:deltaphibound}
    \begin{split}
        \frac{\partial\Delta\phi}{\partial r_c^2}&=-\int_1^\infty\frac{\left(y^2-\frac{1}{y^{d-2}}\right)dy}{2l^2y\left(\frac{r_c^2}{l^2}\left(y^2-\frac{1}{y^{d-2}}\right)+\frac{1}{y^{d-2}}-1\right)^{3/2}}\\
        &<0\\
        \implies  \frac{\partial\Delta\phi}{\partial \mu}&= \frac{\partial\Delta\phi}{\partial r_c^2} \frac{\partial r_c^2}{\partial \mu}\\
        &>0\\
        \implies\Delta\phi&>\Delta\phi|_{\mu=0}\\
        &=\int_{l}^\infty\frac{dr}{r\sqrt{\frac{r^2}{l^2}-1}}\\
        &=\frac{\pi}{2}
    \end{split}
\end{equation}

\begin{figure} 
    \centering
    \includegraphics[scale=0.3]{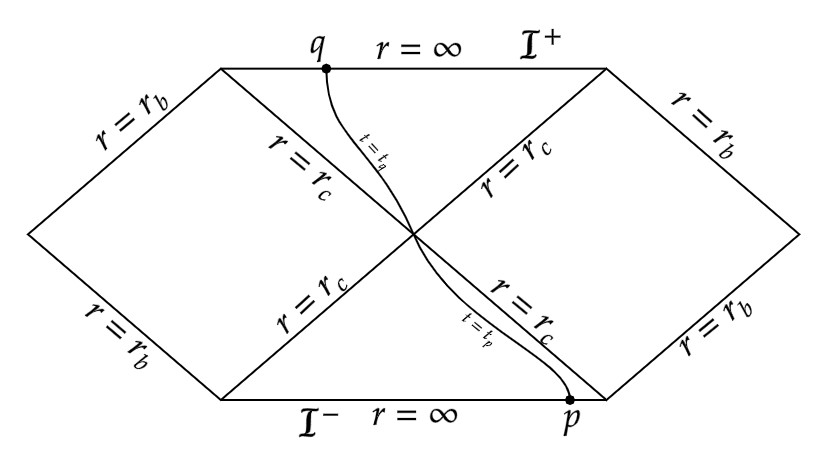}
    \caption{The projection onto the $r-t$ plane of a timelike path from $p$ to $q$ is shown We follow two separate curves of constant $t$. These curves join at $r=r_c$ and can be smoothed to form a path from $p$ to $q$ which is timelike everywhere.}
    \label{fig:SdSpath}
\end{figure}

Using this fact we can construct a timelike path between any two points in $\mathcal{I}^-$ and $\mathcal{I}^+$ using the procedure described below and illustrated in Figure \ref{fig:SdSpath}. 

Choose co-ordinates such that $p$ and $q$ both lie in the equatorial plane and $\phi=0$ at $p$, $\phi=\phi_q\in[0,\pi]$ at $q$. Let $t=t_p$ at $p$ and $t=t_q$ at $q$. Starting at $p$, we follow a path in the equatorial plane with $t=t_p$ until we reach $r=r_c$. We choose the first part of this path to be null, with $\phi$ varying until we reach $\phi=\phi_q/2$ (the above calculation shows that this occurs before we reach $r=r_c$). After this, we choose $\phi$ to remain constant until we reach $r=r_c$, so this part of the path will be timelike. At this point on the Penrose diagram, all $t=$ constant surfaces intersect. This means we can switch to a similar path in the equatorial plane with $t=t_q$ until we reach $r=\infty$. Again we choose this path to be null initially, with $\phi$ varying until we reach $\phi=\phi_q$. After this we set $\phi$ constant, so the path finishes at $q$. This path can then be smoothed to form a timelike curve from $p$ to $q$ \cite{DiffTop}. \qedsymbol

This result should be compared to the results obtained in \cite{PCMD} and \cite{Penrose} (summarised in Theorem A), where it was found that the positive mass asymptotically flat Schwarzschild spacetime satisfies the non-timelike boundary version of the Penrose property in $3+1$ dimensions but not in higher dimensions.

As in the asymptotically flat case, the non-timelike boundary version of the Penrose property fails for negative mass Schwarzschild-de Sitter spacetime. 
.
\begin{thm}\label{thm:negativemassSdS}
The non-timelike boundary version of the Penrose property does not hold for the Schwarzschild-de Sitter spacetime with mass parameter $\mu\leq0$ in $d+1$ dimensions for any $d\geq3$.
\end{thm}
\textbf{Proof:} If $\mu=0$ then this is de Sitter spacetime which was covered in Theorem \ref{thm:dSPenroseproperty}.

\begin{figure}
        \centering
        \includegraphics[scale=0.3]{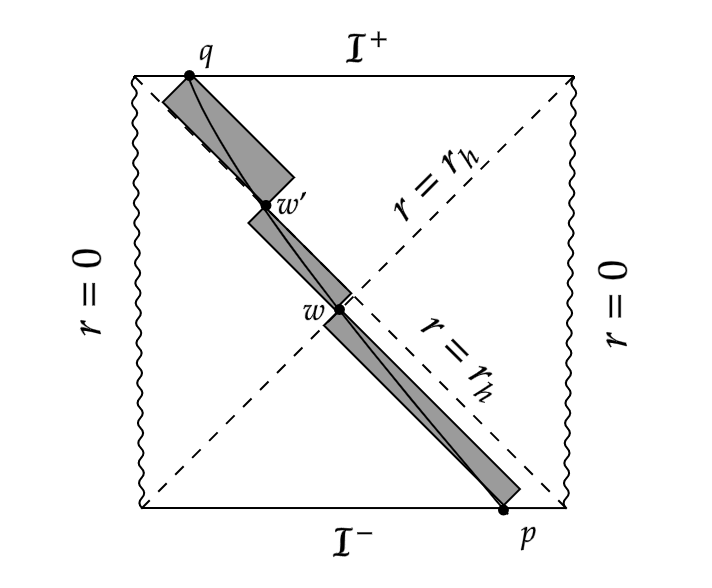}
        \caption{Penrose diagram for the Schwarzschild-de Sitter spacetime with negative mass. This figure shows the construction used in the proof of Theorem \ref{thm:negativemassSdS}.}
        \label{fig:negativemassSdS}
    \end{figure}

Let $\mu<0$. Suppose $p$ and $q$ can be connected by a timelike curve $\gamma$ as shown in Figure \ref{fig:negativemassSdS} and choose co-ordinates so that these points lie in the equatorial plane. Let $w$, $w'$ denote the points on $\gamma$ at which $r=r_h$, where this is the unique root of $V(r)$ (note that $w$ and $w'$ may be the same point). Then since the sets $J^+(p)\cap J^-(w)$, $J^+(w)\cap J^-(w')$ and $J^+(w')\cap J^-(q)$ (shown as the shaded regions in Figure \ref{fig:negativemassSdS}) are non-empty, compact and strongly causal we can replace $\gamma$ with separate causal geodesics from $p$ to $w$, from $w$ to $w'$ and from $w'$ to $q$ \cite[Theorem 1]{Seifert}.

From the Euler-Lagrange equations, we find that along a causal geodesic in the equatorial plane from $r=r_h$ to $r=\infty$, we have (choosing $\phi$ increasing without loss of generality)
\begin{equation}
    \Delta\phi=\int_{r_h}^\infty\frac{h\text{d}r}{r^2\left(E^2-V(r)\left(\frac{h^2}{r^2}+l^2\right)\right)^{1/2}}
\end{equation}
where $E=-V(r)\dot{t}$, $h=r^2\dot{\phi}$, $l^2=\left(\frac{ds}{d\tau}\right)^2$ are constants and $\dot{}$ denotes differentiation with respect to an affine parameter $\tau$. It is clear that $\Delta\phi$ is maximised by setting $E=l=0$. This corresponds to a null geodesic along a line $t=$ constant. These curves were considered in the proof of Theorem \ref{thm:SchwarzschilddeSitter} and here we will denote the change in the co-latitude $\phi$ along such a curve by $\Delta\phi_{max}$. We see from equation (\ref{eqn:deltaphibound}) that
\begin{equation}
\begin{split}
\frac{\partial\Delta\phi_{max}}{\partial\mu}<0\\
\implies\Delta\phi\leq\Delta\phi_{max}<\Delta\phi_{max}|_{\mu=0}&=\frac{\pi}{2}
\end{split}
\end{equation}
So we have an upper bound, $2\Delta\phi_{max}$, on the change in $\phi$ along the curve $\gamma$ excluding the portion between $w$ and $w'$. This upper bound is strictly less than $\pi$ and crucially does not depend on the positions of $p$ and $q$ (a fact which follows from the invariance of the metric under $t-$translations). It is clear that as $p$ and $q$ slide along $\mathcal{I}^-$ and $\mathcal{I}^+$ respectively towards opposite corners of the Penrose diagram (Figure \ref{fig:negativemassSdS}), the geodesic between $w$ and $w'$ approaches a radial geodesic and hence the change in $\phi$ along it tends to 0. It follows that if $p$ and $q$ are chosen to be sufficiently close to opposite corners in the Penrose diagram then the change in $\phi$ along $\gamma$ must be strictly less than $\pi$. If we also choose $p$ and $q$ to be antipodal then we conclude that they cannot be timelike connected. \qedsymbol

\begin{cor}\label{cor:finitePPSdS}
The finite version of the Penrose property holds in Schwarzschild-de Sitter spacetime if and only if the mass parameter is $\mu>0$.
\end{cor}
\textbf{Proof:} This follows immediately from Theorems \ref{thm:dsequivalence}, \ref{thm:SchwarzschilddeSitter} and \ref{thm:negativemassSdS}. \qedsymbol

Note that if we were instead to allow curves which did not escape to the asymptotic region (and for consistency extend the non-timelike boundary version of the Penrose property to include their endpoints at $i^\pm$) then Theorem \ref{thm:SchwarzschilddeSitter} and Corollary \ref{cor:finitePPSdS} would no longer be true. In particular, we would be able to choose endless timelike curves remaining in the two causally disconnected regions between the event and cosmological horizons (see Figure \ref{fig:SchwarzschilddeSitter}). The past endpoint at $i^-$ of one of these curves would be timelike disconnected from the future endpoint at $i^+$ of the other. The complication here arises because the conformal boundary is a spacelike hypersurface. As a result it is possible for the non-timelike boundary version of the Penrose property as stated in Definition \ref{defn:Penroseproperty} to hold and yet for there to be points at past and future timelike infinity which are not timelike connected.

The results of this section provide justification for our choice to exclude timelike curves which fail to reach the asymptotic region. Recall that Penrose's original motivation for studying the Penrose property was to make comparisons between the spacetime of interest and an appropriate ``background'' spacetime. In the case of \cite{Penrose} this background was Minkowski spacetime, while in this section the relevant background is de Sitter. Our conventions give us a property which is strong enough that when we find a physically relevant spacetime which satisfies it, Theorem \ref{thm:backgrounddS} ensures that this is enough to rule out any ``$SO(d+1,1)$ covariant'' construction of quantum gravity based on a background de Sitter spacetime. In light of Theorem \ref{thm:SchwarzschilddeSitter}, we remark that the positive mass Schwarzschild-de Sitter spacetime provides such an example. If we were to include the points $i^\pm$ in the definition of the non-timelike boundary version of the Penrose property then this property would not be satisfied by the positive mass Schwarzschild de-Sitter spacetime and we would be unable to use Theorem \ref{thm:backgrounddS} to rule out such quantum gravity constructions. 

\subsection{The Penrose Property in Asymptotically Anti-de Sitter Spacetime}\label{The Penrose Property in Asymptotically Anti-de Sitter Spacetime}
Next we consider spacetimes which are asymptotically anti-de Sitter. The boundary of the conformal compactification, denoted $\partial \overline{M}$, is timelike and cannot be separated into ``past'' and ``future'' components. As a result it is not obvious what the analogue of the non-timelike boundary version of the Penrose property should be. 

To address this, we begin by considering anti-de Sitter spacetime in $d+1$ dimensions. This spacetime is an isometrically embedded submanifold of $\text{Mink}_{d,2}$ defined by
\begin{equation}\label{eqn:adssubmanifold}
    -x_0^2-x_1^2+\sum_{i=2}^{d+1}x_i^2=-l^2
\end{equation}
where $(x_0,x_1,...,x_{d+1})$ are a standard Cartesian co-ordinate system on $\mathbbm{R}^{d+1,1}$ and the radius of curvature, $l$, is a positive constant which is related to the cosmological constant by
\begin{equation}
    \Lambda=-\frac{d(d-1)}{2l^2}
\end{equation}

The metric in static co-ordinates is given by 
\begin{equation}
    ds^2=V(r)dt^2-\frac{dr^2}{V(r)}-r^2d\omega^2_{d-1}
\end{equation}
where $V(r)=1+r^2/l^2$.

In order to formulate an appropriate analogue of the non-timelike boundary version of the Penrose property, we begin with a result from \cite{HorowitzItzhaki}. Given $p\in\partial \overline{M}$, it will be useful to distinguish between the set of points in $\partial \overline{M}$ which can be reached by future pointing causal curves from $p$ which either remain on $\partial \overline{M}$ or which have only their endpoints on $\partial \overline{M}$. We define the following sets
\begin{equation}
\begin{split}
    A(p)&=\{q\in\partial \overline{M}:\exists \text{ future directed causal curve } \gamma \text{ from } p\text{ to }q \text{ such that } \gamma\setminus\{p,q\}\subset \text{int}(\overline{M})\}\\
    B(p)&=\{q\in\partial \overline{M}:\exists \text{ future directed causal curve } \gamma \text{ from } p\text{ to }q \text{ such that } \gamma\subset\partial \overline{M}\}
    \end{split}
\end{equation}
where the interior of a set $X$, denoted $\text{int}(X)$, is defined to be the union of all open subsets of $X$.

We will also make use of the fact that the timelike future of a point is an open set, and hence
\begin{equation}
    \text{int}(A(p))=\{q\in\partial \overline{M}:\exists \text{ future directed timelike curve } \gamma \text{ from } p\text{ to }q \text{ such that } \gamma\setminus\{p,q\}\subset \text{int}(\overline{M})\}
\end{equation}

In order to define an appropriate boundary version of the Penrose property, the following result will be useful.

\begin{thm}[Horowitz, Itzhaki \cite{HorowitzItzhaki}]\label{thm:ads}
$A(p)=B(p)$ in Anti-de Sitter spacetime in $d+1$ dimensions for any $d\geq2$.
\end{thm}
\begin{figure} 
    \centering
    \includegraphics[scale=0.2]{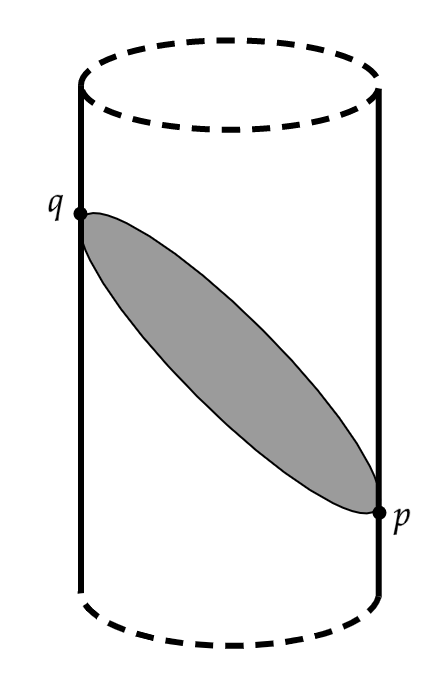}
    \caption{Diagram showing a portion of conformal anti-de Sitter spacetime with its timelike boundary. The shaded region is the future lightcone of a point $p$ on this boundary. All future pointing null geodesics from $p$ (including those restricted to the boundary) have the same future endpoint at $q$.}
    \label{fig:AdSlightcone}
\end{figure}

\textbf{Proof}: We write the metric in Poincar\'{e} co-ordinates
\begin{equation}\label{eqn:Poincare}
    ds^2=\frac{l^2}{z^2}(dt^2-dz^2-\sum_{i=1}^{d-1}(dx^i)^2)
\end{equation}
where $0\leq z<\infty$, with $z=0$ corresponding to points on the conformal boundary. The disadvantage of these co-ordinates is that they only cover part of the spacetime (the ``Poincar\'{e} patch''). The advantage is that they show the AdS metric to be conformally flat. The conformally compactified metric is
\begin{equation}\label{eqn:conformalads}
    \overline{ds^2}=dt^2-dz^2-\sum_{i=1}^{d-1}(dx^i)^2
\end{equation}

Consider a null plane in the compactified spacetime. On this plane we can choose a sequence of curves approaching the boundary so that, when we parameterize these curves by an affine parameter $\tau$, the sequence of functions $z(\tau)\rightarrow0$ uniformly. But from equation (\ref{eqn:conformalads}), this means that $\dot{\tau}^2-\sum_{i=1}^{d-1}(\dot{x}^i)^2\rightarrow0$ uniformly along this sequence (where $\dot{}$ denotes differentiation with respect to $\tau$), and hence the limit curve on the boundary $z=0$ is null. 

This tells us that the intersection between the conformal boundary and a null plane in the compactified spacetime is a null curve\footnote{This is in contrast with the intersection bewteen a null plane and a timelike cylinder in Minkowski spacetime which gives a spacelike curve.}. We conclude that $A(p)=B(p)$. \qedsymbol

The future null cone of a point $p$ on the conformal boundary is shown in Figure \ref{fig:AdSlightcone}.

With this result in mind, we define a boundary version of the Penrose property in asymptotically AdS spacetimes as follows:
\begin{definition}[Penrose property - timelike boundary version]\label{defn:PenrosePropertyAdS}
Let $(M,g)$ be an asymptotically AdS spacetme with compactification $(\overline{M},\overline{g})$. Then $(M,g)$ satisfies the \textit{timelike boundary version of the Penrose property} if $B(p)\subset \text{int}(A(p))$ for any $p\in\partial \overline{M}$.
\end{definition}
This definition is similar to the one used for asymptotically flat spacetimes in the following sense. First note that it is not satisfied in AdS spacetime, however from Theorem \ref{thm:ads} we see that it is ``almost'' satisfied. Similarly, the non-timelike boundary version of the Penrose property ``almost'' holds in Minkowski spacetime, where only antipodal points on $\mathcal{I}^-$ and $\mathcal{I}^+$ lying sufficiently close to $i^0$ cannot be timelike connected. The only null curve between these points is the one lying entirely on the conformal boundary and passing through $i^0$. This is shown in Figure \ref{fig:spatialinfinity}. The non-timelike boundary version of the Penrose property for an asymptotically flat spacetime (Definition \ref{defn:Penroseproperty}) asks whether the curvature of this spacetime allows us to do better and connect such points (along with all other pairs of points on $\mathcal{I}^-$ and $\mathcal{I}^+$) using a timelike curve through the interior of the spacetime. Similarly, our definition of the timelike boundary version of the Penrose property asks if the curvature of the spacetime allows us to use timelike curves through the interior to connect points which can only be connected by null curves on the Minkowski conformal boundary\footnote{There is a subtlety here because asymptotically flat spacetimes with non-zero ADM mass will be singular at $i^0$. For this reason we would need to consider the conformal boundary as a spacetime in its own right on which the metric is conformal to Minkowski. In this case a point can be inserted at $i^0$ and we can connect antipodal points using null curves through this point.}.

We stress that, as in the asymptotically flat case, this property is a property of spacetimes near the conformal boundary. In particular, suppose we can find a timelike curve through the interior spacetime from $p$ to $q$, for some $q\in B(p)$. Then we can also find such a curve from $p$ to $w$ for any $w\in B(q)$. This is done by smoothing the timelike curve from $p$ to $q$ and the boundary causal curve from $q$ to $w$ in such a way as to get a timelike curve through the interior with the same endpoints. But by choosing $q$ arbitrarily close to $p$, any timelike curve between them must remain arbitrarily close to the boundary and its existence is determined by the asymptotic properties of the metric. This is in contrast to the behaviour observed in the asymptotically de Sitter case, where it was necessary to consider curves which leave the asymptotic region. As a result, the non-timelike boundary version of the Penrose property in asymptotically de Sitter spacetimes would appear to be more complicated than the same version in asymptotically flat spacetimes or the timelike boundary version in asymptotically AdS spacetimes. 

\begin{figure}
    \centering
    \includegraphics[scale=0.3]{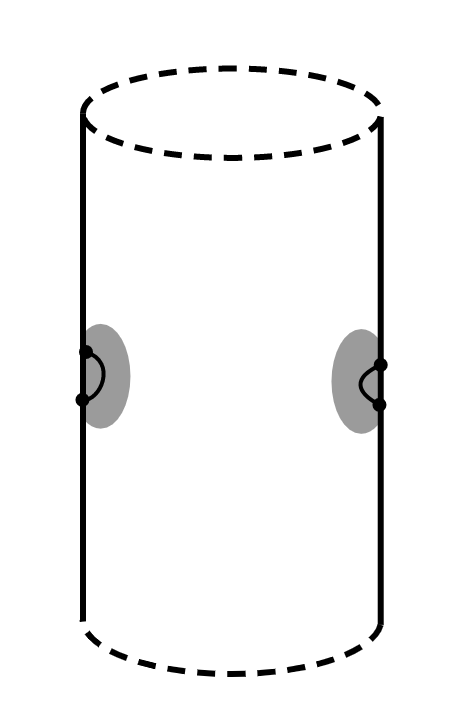}
    \caption{This figure shows how in an asymptotically AdS spacetime we can choose two timelike curves (endless in the uncompactified spacetime) with endpoints in open sets on the timelike boundary which are not timelike connected.}
    \label{fig:finitePenroseAdsfails}
\end{figure}

We now state the result of Gao and Wald which shows that the timelike boundary version of the Penrose property fails in spacetimes which focus null geodesics entering the bulk.

\begin{thm}[Gao, Wald \cite{GW}]\label{thm:GaoWald}
Let $(M,g)$ be a spacetime with conformal compactification $(\overline{M},\overline{g})$ such that
\begin{enumerate}
    \item Every complete null geodesic in $(M,g)$ contains a pair of conjugate points.\footnote{To guarantee this, we could impose the null generic condition and the null energy condition, although this latter condition could be replaced by the weaker averaged null energy condition due to Borde \cite{Borde}.}
    \item $\text{int}(\overline{M})$ is strongly causal
    \item For any $p,q\in\text{int}(\overline{M})$, $J^+(p)\cap J^-(q)$ is compact
    \item $\partial \overline{M}$ is a timelike hypersurface in $\overline{M}$.
\end{enumerate}

Then $\partial A(p)\subset B(p)$, so in particular the timelike boundary version of the Penrose property fails.
\end{thm}
\textbf{Sketch of Proof (following \cite{GW}):} We begin by showing that $A(p)$ ($p\in\partial\overline{M}$) is open. Suppose $r\in A(p)$ and let $\lambda$ be a causal curve from $p$ to $r$ which otherwise does not intersect $\partial\overline{M}$. If $\lambda$ is not a null geodesic then it can be deformed to become a timelike curve with the same endpoints. If $\lambda$ is a null geodesic then it contains a pair of conjugate points (assumption 1) and can be similarly deformed to a timelike curve without changing its endpoints. This shows that $r\in I^+(p)$ and hence there exists a neighbourhood of $r$ contained in $A(p)$ We conclude that $A(p)$ is open.

Suppose $p,q\in\partial\overline{M}$ and $q\in\partial A(p)$. This tells us that $q\notin I^+(p)$, otherwise an open neighbourhood of $q$ would be contained in $A(p)$. By taking the limit of a sequence of timelike curves from $p$ with endpoints approaching $q$ (and using the fact that $J^+(p)\cap J^-(q)$ is compact and $\overline{M}$ is timelike and strongly causal) we can construct a causal curve, $\gamma$, from $p$ to $q$ (see \cite{GW} for details of this construction). Now if a segment of $\gamma$ enters the interior of the spacetime, then by the above argument this segment can be deformed if necessary to become timelike (without changing its endpoints). As a result, the entire curve $\gamma$ can be deformed to give a timelike curve from $p$ to $q$. This is a contradiction, so we conclude that $\gamma$ lies entirely in $\partial\overline{M}$ and hence $q\in B(p)$. \qedsymbol

As explained in \cite{GW}, Theorem \ref{thm:GaoWald} says that null geodesics through spacetimes satisfying the conditions of the theorem are ``delayed'' relative to null geodesics in pure AdS. To make this comparison we use null geodesics on the conformal boundary as a reference. This is actually the same effect as is observed in positive mass asymptotically flat spacetimes and is the reason that the non-timelike boundary version of the Penrose property can be shown to fail in higher dimensional positive mass Schwarzschild (see \cite[Theorem 4.2]{PCMD} and the discussion in Section \ref{Introduction}). 

From Theorem \ref{thm:ads}, we immediately get the following analogue of Theorem IV.5 in \cite{Penrose}.
\begin{thm}\label{thm:AdSbackground}
Let $(\overline{M},\overline{g})$ be the conformal compactification of an asymptotically anti-de Sitter spacetime which satisfies the timelike boundary version of the Penrose property and let $\mathcal{I}$ denote its conformal boundary at infinity. Then, given any $p\in\partial\overline{M}$, there is no compactification of AdS spacetime which has $\mathcal{I}$ as its boundary and $\overline{g}\leq\overline{g}_{AdS}$ on some neighbourhood of $p$ in $\overline{M}$.
\end{thm}
\textbf{Proof:} Suppose such a compactification exists and let $p\in\partial\overline{M}$ be a point with neighbourhood $U\subset\overline{M}$ on which $\overline{g}\leq\overline{g}_{AdS}$ holds. Let $q\in B_0(p)$. Since $B(p)=B_0(p)$, there exists a curve $\lambda\subset\partial\overline{M}$  from $p$ to $q$ which is causal with respect to $\overline{g}$. Since $(\overline{M},\overline{g})$ satisfies the timelike version of the Penrose property, given any $p'\in\lambda$ there exists a $\overline{g}$--causal curve through the interior spacetime from $p$ to $p'$. By choosing $p'$ sufficiently close to $p$, this curve must lie entirely in $U$.  The condition $\overline{g}\leq\overline{g}_{AdS}$ implies that this curve is causal with respect to the metric $\overline{g}_{AdS}$. We therefore have a piecewise $\overline{g}_{AdS}$-causal curve from $p$ to $q$ which lies in the interior from $p$ to $p'$ and on the boundary from $p'$ to $q$. Since $\partial\overline{M}$ is totally geodesic as a submanifold of $\overline{M}$, this curve is not a null geodesic, so can be smoothed to give a $\overline{g}_{AdS}$ timelike curve from $p$ to $q$ which lies entirely in the interior spacetime \cite{DiffTop}. It follows that $q\in\text{int}(A_0(p))$. We conclude that $B_0(p)\subset\text{int}(A_0(p))$ and hence $B_0(p)\neq A_0(p)$, since both of these sets are closed. This contradicts Theorem \ref{thm:ads}. \qedsymbol


\subsection{Schwarzschild-Anti de Sitter Spacetime}\label{Schwarzschild-anti de Sitter Spacetime}
As an example and to consider the similarities with the results of both Section \ref{Schwarzschild-de Sitter} and of \cite{PCMD}, we consider the Schwarzschild-anti de Sitter spacetime. In $d+1$ dimensions, the metric for this spacetime is 
\begin{equation}\label{eqn:Schwarzschild-AdS}
\begin{split}
    ds^2&=V(r)dt^2-\frac{dr^2}{V(r)}-r^2d\omega_{d-1}^2\\
    V(r)&=1-\frac{\mu}{r^{d-2}}+\frac{r^2}{l^2}
    \end{split}
\end{equation}
where we have once again introduced a \textit{mass parameter}, $\mu$, which is related to the mass of the AdS ground state \cite{WittenAdS}, $m$, by 
\begin{equation}
    \mu=\frac{16\pi m}{(d-1)A_{S^{d-1}}}
\end{equation}

\begin{figure}
    \centering
    \includegraphics[scale=0.3]{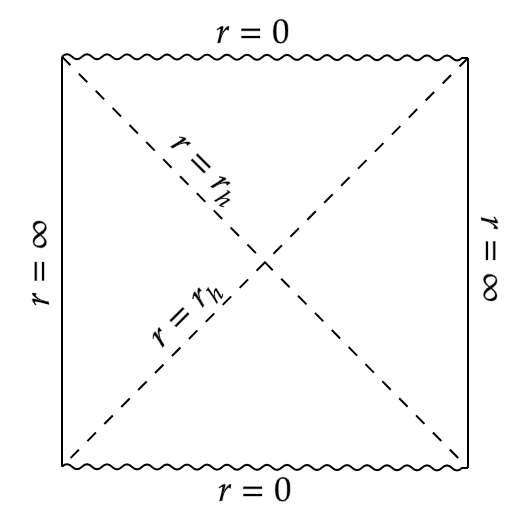}
    \caption{Penrose diagram for the positive mass Schwarzschild-AdS spacetime. This spacetime contains a spacelike singularity at $r=0$, an event horizon at $r=r_h$ (where $V(r_h)=0$) and a timelike conformal boundary.}
    \label{fig:SchwarzschildAdS}
\end{figure}

The Penrose diagram for the positive mass Schwarzschild-AdS spacetime is shown in Figure \ref{fig:SchwarzschildAdS}. This figure also shows how the finite version of the Penrose property can be seen to fail in this spacetime. 

\begin{thm}\label{thm:SchwarzschildAdS}
The Schwarzschild-AdS spacetime in $d+1$ dimensions ($d\geq3$) satisfies the timelike boundary version of the Penrose property if and only if the mass parameter $\mu<0$.
\end{thm}
Note that the Schwarzschild-AdS spacetime does not satisfy the conditions of Theorem \ref{thm:GaoWald} so is not covered by this result. In particular, if $\mu<0$ then there is a naked singularity and condition 3 is not satisfied. If $\mu>0$ then the endless null geodesic along the horizon does not contain any conjugate points (indeed the null generic conditions fails for this geodesic). If $\mu=0$ then any radial null geodesic has no conjugate points (once again the null generic condition fails). This explains why Theorem \ref{thm:ads} does not contradict Theorem \ref{thm:GaoWald}.

\textbf{Proof}: We begin by considering the case where $\mu<0$ and we aim to show that $B(p)\subset \text{int}(A(p))$. In this case $V(r)>0$ for all $r>0$, so the singularity at $r=0$ is naked.

We first compactify the Schwarzschild-AdS metric using the same procedure as was used in \cite[Section 4.2.1]{PCMD}. We define the tortoise co-ordinate, $r_*$, by
\begin{equation}
    \begin{split}
        dr_*&=\frac{dr}{V(r)}\\
        &=\frac{dr}{1+\frac{r^2}{l^2}}\left(1+O(r^{-d})\right)\\
        \implies r_*&=l\tan^{-1}(r/l)+O(r^{-(d-1)})
    \end{split}
\end{equation}

We see that all terms involving $\mu$ are sub-leading, and in particular tend to $0$ as $r\rightarrow\infty$. From this it is clear that the co-ordinates used to compactify the Schwarzschild-AdS metric (equation (\ref{eqn:Schwarzschild-AdS})) agree at $r=\infty$ with those used to compactify pure anti-de Sitter spacetime. This allows us to use a spacetime comparison argument similar to those used in \cite{PCMD}\footnote{See footnote 5 in \cite{PCMD} for a discussion of why the agreement of compactified co-ordinates on the conformal boundary is crucial for making spacetime comparisons such as the one used in this proof.} and \cite{HeWuXie}. 

In what follows we use the subscript 0 to denote pure AdS spacetime (i.e. Schwarzschild-AdS with $\mu=0$) with the same radius of curvature, $l$, as the Schwarzschild-AdS spacetime we are considering. We have  
\begin{equation}\label{eqn:inequality}
    \begin{split}
        V(r)&> V_0(r)\\
        \implies ds^2&>ds_0^2\\
        \implies \overline{ds}^2&>\frac{\Omega^2_0}{\Omega^2}\overline{ds}_0^2
    \end{split}
\end{equation}
where $\Omega^2$ and $\Omega_0^2$ are the conformal compactification factors for the Schwarzschild-AdS and pure AdS spacetimes respectively. These can be obtained using the same procedure as was used in \cite[Section 4.2.1]{PCMD} to compactify the asymptotically flat Schwarzschild spacetime. Recall that these factors are strictly positive on the interior spacetime and that the two metrics agree on the boundary. As a result, the set $B(p)$ is the same for both metrics. We denote this $B(p)=B_0(p)$.

Now suppose $p$, $q\in\partial \overline{M}$ and $q\in B(p)$. Then $q\in B_0(p)=A_0(p)$, so $p$ and $q$ are connected by a curve lying in the interior spacetime (other than its endpoints) which is causal with respect to the compactified AdS metric. From inequality (\ref{eqn:inequality}), we conclude that this curve is timelike with respect to the compactified negative mass Schwarzschild-AdS metric. We conclude that $q\in \text{int}(A(p))$.

If $\mu>0$ then a similar argument shows that the timelike boundary version of the Penrose property does not hold. Note that in this case the equation $V(r)=0$ has exactly one positive root corresponding to the event horizon. However, since we are interested in timelike connected points on the conformal boundary, we can restrict attention to regions where $r$ is large and hence $V(r)>0$ (see the comment in Section \ref{The Penrose Property in Asymptotically Anti-de Sitter Spacetime} explaining why the timelike boundary version of the Penrose property is a property of spacetimes near their timelike boundary). We find that inequality (\ref{eqn:inequality}) is reversed, so we have 
\begin{equation}\label{eqn:inequalityreverse}
    \begin{split}
        \overline{ds}^2<\frac{\Omega^2_0}{\Omega^2}\overline{ds}_0^2
    \end{split}
\end{equation}
Now suppose $p$, $q\in\partial \overline{M}$ and $q\in\partial B(p)$. Suppose $q\in \text{int}(A(p))$, so $p$ and $q$ are connected by a curve through the interior spacetime which is timelike with respect to the compactified Schwarzschild-AdS metric. Inequality (\ref{eqn:inequalityreverse}) implies that this curve is also timelike with respect to the compactified AdS metric, so $q\in \text{int}(A_0(p))$ and hence $q\in \text{int}(B_0(p))$ by Theorem \ref{thm:ads}. But recall the AdS and AdS-Schwarzschild metrics agree on the conformal boundary, so $\text{int}(B_0(p))=\text{int}(B(p))$ and hence $q\in \text{int}(B(p))$. This is a contradiction. \qedsymbol

It turns out that the finite version of the Penrose property (Definition \ref{defn:Penrosefinite}) does not generalise to an interesting property in asymptotically AdS spacetimes.
\begin{thm}
For asymptotically anti-de Sitter spacetimes, the timelike boundary version of the Penrose property is not equivalent to the finite version of the property which fails trivially. 
\end{thm}
\textbf{Proof: }Since the conformal boundary of an asymptotically AdS spacetime is timelike, we can find two open sets on this boundary which are not timelike connected. We then choose two causal curves (endless in the uncompactified spacetime) one with both endpoints in the first of these open sets and the other with both endpoints in the second. Since the open sets are not timelike connected it follows that neither are these two curves. This is illustrated in Figure \ref{fig:finitePenroseAdsfails}.

From Theorems \ref{thm:ads} and \ref{thm:SchwarzschildAdS}, we see that there are examples of asymptotically AdS spacetimes which do and do not satisfy the timelike boundary version of the Penrose property, so we conclude that this property is not equivalent to the finite version. \qed

Theorem \ref{thm:SchwarzschildAdS} should be contrasted with the results obtained for the Schwarzschild-de Sitter spacetime and for the 3+1 dimensional asymptotically flat Schwarzschild spacetime. This is summarised by Theorem B in Section \ref{Introduction}. It seems that for asymptotically AdS spacetimes, the timelike boundary version of the Penrose property is a property of spacetimes which negative mass. This agrees with Theorem \ref{thm:GaoWald} which associates the failure of this property with spacetimes which focus null geodesics, a property we associate with global positivity of mass via the positive mass theorem. 

If we regard the negative mass AdS-Schwarzschild spacetime as unphysical, then Theorem \ref{thm:SchwarzschildAdS} means that we are unable to use Theorem \ref{thm:AdSbackground} to rule out a ``$SO(d,2)$ covariant'' construction of quantum gravity defined with respect to a fixed background anti-de Sitter spacetime. 

\section{Further Examples of the Penrose Property in Asymptotically Flat Spacetimes}
We now move on to consider some additional example spacetimes, each exhibiting a feature which was not encountered in \cite{PCMD}. We will discuss the relevance of these features for the non-timelike boundary version of the Penrose property (Definition \ref{defn:Penroseproperty}), which will be equivalent to the finite version (Definition \ref{defn:Penrosefinite}) in each case.

\subsection{Product Spacetimes}\label{Product Spacetimes}
In this section we consider spacetimes of the form $(M,g)=(M',g')\times(M'',g'')$, where $(M',g')$ is an asymptotically flat Lorentzian manifold and $(M'',g'')$ is a compact Riemannian manifold. Before proceeding we must consider how to compactify such spacetimes. We can conformally embed $M$ into the warped product manifold $\tilde{M}=\tilde{M}'\times_{\Omega'} M''$ endowed with the metric
\begin{equation}
\overline{g}:=\Omega'^2g=\overline{g}'-\Omega'^2 g''
\end{equation}
where $\overline{g}'=\Omega'^2g'$ and we have omitted the embedding map for brevity\footnote{Recall that we are using the `mostly minus' signature $(+,-,...,-)$.}. 

Since $\Omega'=0$ on $\partial\overline{M}'$, the metric $\overline{g}$ degenerates to $\overline{g}'$ on $\partial\overline{M}'\times M''$ and we have conical singularities. For this reason, we define
\begin{equation}
\begin{split}
    \partial\overline{M}&=\left(\left(\partial\overline{M}'\cup i^-\cup i^+\right)\times M''\right)/\sim\\
    &\cong\partial\overline{M}'
\end{split}
\end{equation}
where the equivalence relation $\sim$ is defined by
\begin{equation}
    (x,y_1)\sim (x,y_2) \text{ for any $x\in\partial\overline{M}'$, $y_1,y_2\in M''$}.
\end{equation}
We then define $\overline{M}:=(\text{int}(\overline{M}')\times M'')\cup\partial\overline{M}$ and refer to $(\overline{M},\overline{g})$ as the conformal compactification of $(M,g)$. The addition of $\partial\overline{M}$ causally completes $M$ (endless causal curves in $M$ have endpoints on $\partial\overline{M}$). Despite this, $\overline{M}$ is not a manifold with boundary, since $\partial\overline{M}$ is not co-dimension 1 in $\overline{M}$. However, it still makes sense to consider the non-timelike boundary version of the Penrose property for such product spacetimes, where the set $\partial\overline{M}$ inherits the splitting of $\partial\overline{M}'$ into ``past'' and ``future'' components, denoted $\mathcal{I}^-$ and $\mathcal{I}^+$ respectively.

\begin{thm}\label{thm:productconformal}
The product spacetime $(M,g)=(M',g')\times(M'',g'')$, where $(M',g')$ is an asymptotically flat Lorentzian manifold and $(M'',g'')$ is a compact Riemannian manifold, satisfies the non-timelike boundary version of the Penrose property if and only if this property is satisfied in $(M',g')$.
\end{thm}
\textbf{Proof:} First note that if we modify any timelike curve in $(\overline{M},\overline{g})$ to have zero length in $(\overline{M}'',\overline{g}'')$ while keeping its path in $(\overline{M}',\overline{g}')$ unchanged then this curve remains timelike. Moreover, if the original curve had endpoints on $\partial\overline{M}$, then since $M''$ does not contribute to $\partial\overline{M}$, our modified curve has the same endpoints as the original. By making such modifications and identifying the resulting curve in $(\overline{M},\overline{g})$ with a curve in $(\overline{M}',\overline{g}')$ in the obvious manner, we see that the non-timelike boundary version of the Penrose property in $(\overline{M},\overline{g})$ is satisfied if and only if it is satisfied in $(\overline{M}',\overline{g}')$. \qedsymbol

In fact, the finite version of the Penrose property in $(M,g)$ is also equivalent to this same property in $(M',g')$.

\begin{thm}\label{thm:productfinite}
The product spacetime $(M,g)=(M',g')\times(M'',g'')$, where $(M',g')$ is an asymptotically flat Lorentzian manifold and $(M'',g'')$ is a compact Riemannian manifold, satisfies the finite version of the Penrose property if and only if this property is satisfied in $(M',g')$.
\end{thm}
\textbf{Proof:} It is clear that if $(M,g)$ satisfies the finite version of the Penrose property then so does $(M',g')$: we simply project all timelike curves in $(M,g)$ to curves in $(M',g')$ which must also be timelike.

Suppose $(M',g')$ satisfies the finite version of the Penrose property and let $\gamma_1=\gamma_1'\times\gamma''_1$ and $\gamma_2=\gamma_2'\times\gamma''_2$ be endless timelike curves in $(M,g)$. Then there exists $(p',p'')\in\gamma_1$, $(q',q'')\in\gamma_2$ such that $p'$ and $q'$ can be connected in $(M',g')$ by a timelike curve $\gamma'$. 

Define the curve $\gamma\subset M$ as follows. In $M'$ we define it to follow the same path as $\gamma'$ from $p'$ to $q'$ and then the same path as $\gamma'_2$ into the infinite future. In $M''$, we define this curve to have zero length (so it consists of a single point). We then smooth \cite{DiffTop} $\gamma$ to obtain a curve with past endpoint at $(p',p'')$ which is timelike in $(M,g)$ and whose future endpoint in the compactified spacetime $(\overline{M},\overline{g})$ agrees with that of $\gamma_2$. Hence these curves have the same timelike past (recall that $M''$ does not contribute to the boundary of $\overline{M}$, so the paths of $\gamma$ and $\gamma_2$ in $M''$ are irrelevant). This means in particular that $(p',p'')$ lies in the timelike past of $\gamma_2$ and hence there must exist a point $w\in\gamma_2$ such that $(p',p'')$ lies in the timelike past of $w$, as required. \qed

We use Theorems \ref{thm:productconformal} and \ref{thm:productfinite} to consider the example of the black string spacetime.

\begin{example}[The Black String]\label{example:blackstring}
The finite black string metric, defined on $\left(\mathbbm{R}^{d,1}\setminus\{0\}\right)\times S^1$, is given by \cite{EmparanReall}
\begin{equation}\label{eqn:blackstring}
    \begin{split}
        ds^2=-V(r)dt^2+\frac{dr^2}{V(r)}+r^2d\omega_{d-1}^2+dz^2
    \end{split}
\end{equation}
where $V(r)=1-\frac{\mu}{r^{d-2}}$ and the mass parameter, $\mu$, is related to the ADM mass, $m$, by
\begin{equation}
    \mu=\frac{16\pi m}{(d-1)A_{S^{d-1}}}
\end{equation}
We also have $z\sim z+L$ for some constant $L\in(0,\infty)$. 
\begin{figure} 
    \centering
    \includegraphics[scale=0.25]{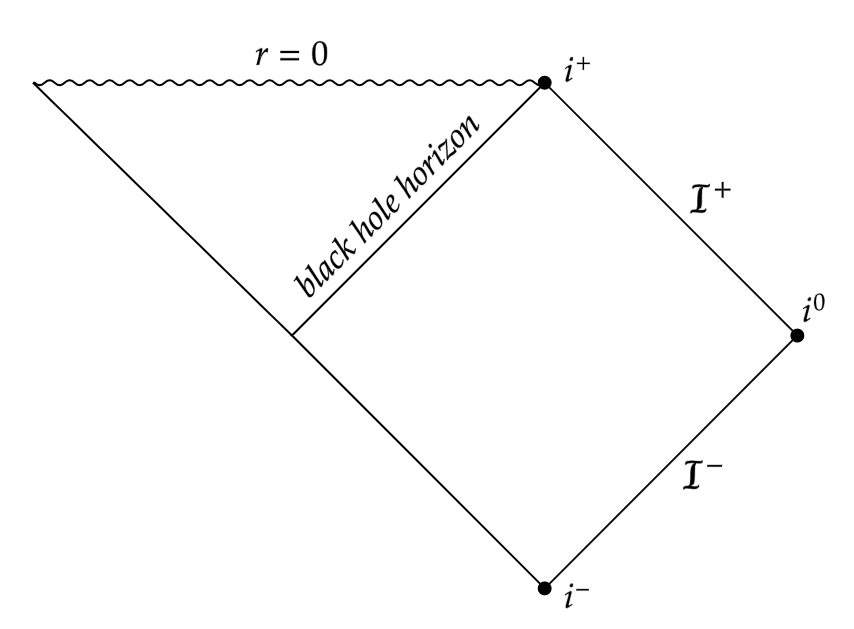}
    \caption{The Penrose diagram for the black string spacetime (shown here in the case $\mu>0$) is the same as for Schwarzschild except each interior point now represents a manifold which is topologically $S^{d-1}\times S^1$ rather than $S^{d-1}$.}
    \label{fig:blackstringpenrose}
\end{figure}

The Penrose diagram for this spacetime (Figure \ref{fig:blackstringpenrose}) is the same as the Penrose diagram for Schwarzschild except each interior point now represents a manifold which is topologically $S^{d-1}\times S^1$ rather than $S^{d-1}$. Note that endless null geodesics through $\overline{M}$ can now have endpoints at timelike infinity. For example, the curve with $r,\phi,\theta_1,...,\theta_{d-2}$ constant (where $\phi,\theta_1,...,\theta_{d-2}$ are the usual spherical polar co-ordinates on $S^{d-1}$) and $\frac{dz}{dt}=\sqrt{V(r)}$ is a null geodesic with endpoints on $i^\pm$. Note that the $z$ co-ordinate along this null geodesic does not approach a constant value in the infinite future or past.

Along with Theorem A, Theorems \ref{thm:productconformal} and \ref{thm:productfinite} tell us that the two equivalent versions of the Penrose property are satisfied by the finite black string spacetime according to the table below
\begin{center}
    \begin{tabular}{|c|c|c|}\hline \vtop{\hbox{\strut Spacetime dimension}\hbox{\strut of Schwarzschild part}}
  & \vtop{\hbox{\strut }\hbox{\strut $m>0$}}& \vtop{\hbox{\strut }\hbox{\strut $m\leq0$}}\\
   \hline
 $4$ & \cmark & \xmark\\  
 $\geq5$ & \xmark &  \xmark\\
 \hline
    \end{tabular}
\end{center}
\end{example}

It is straightforward to extend Corollary \ref{cor:backgroundMink} to instead refer to a background spacetime of
\begin{equation}
    \text{Mink}_{d,1}\times(M'',g'')
\end{equation}
where $(M'',g'')$ is a compact Riemannian manifold. Example \ref{example:blackstring} allows us to rule out a self-consistent quantum gravity construction for $d=3$ (but not for $d\geq4$) based on this background spacetime.

\subsection{Ellis-Bronnikov Wormhole}\label{Ellis-Bronnikov Wormhole}

The Ellis-Bronnikov wormhole spacetime in $(d+1)$-dimensions ($d\geq2$) has metric
\begin{equation}\label{eqn:wormholemetric}
    ds^2=dt^2-dr^2-(r^2+a^2)d\omega_{d-1}^2
\end{equation}
where $a>0$ is a constant.

\begin{figure} 
    \centering
    \includegraphics[scale=0.15]{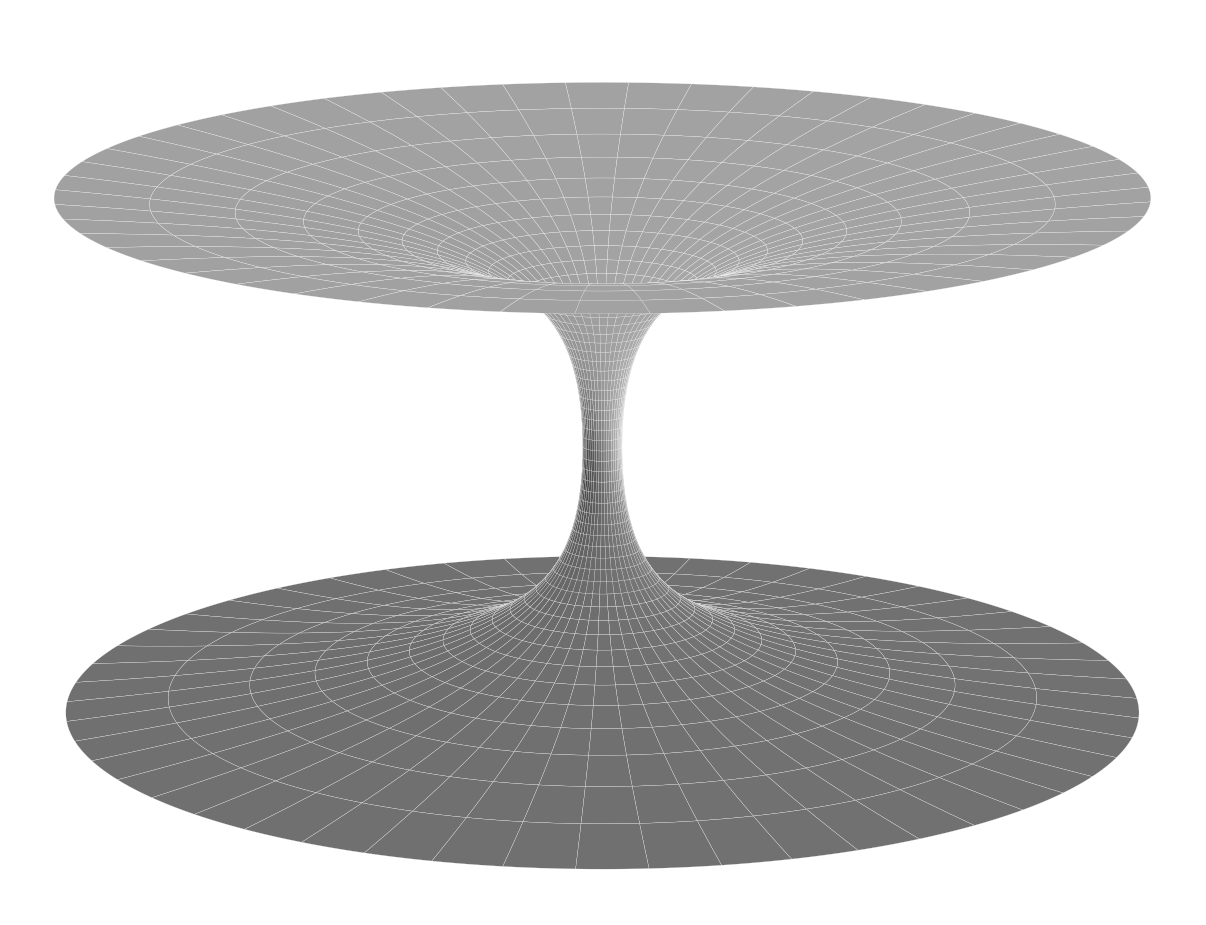}
    \caption{The Ellis-Bronnikov wormhole consists of two asymptotically flat ``universes" separated by a ``throat''. The wormhole is called ``traversable'' because timelike observers can pass through it from one universe to the other. This figure shows the equatorial plane in a $t=$ constant slice of the spacetime, embedded in 3d Euclidean space.}
    \label{fig:wormholethroat}
\end{figure}

In this spacetime, it is possible to extend $r$ through the ``throat" at $r=0$ and consider negative values of $r$. The constant $a$ corresponds to the width of this throat. An illustration of this spacetime is given in Figure  \ref{fig:wormholethroat}.

\begin{figure} 
    \centering
    \includegraphics[scale=0.3]{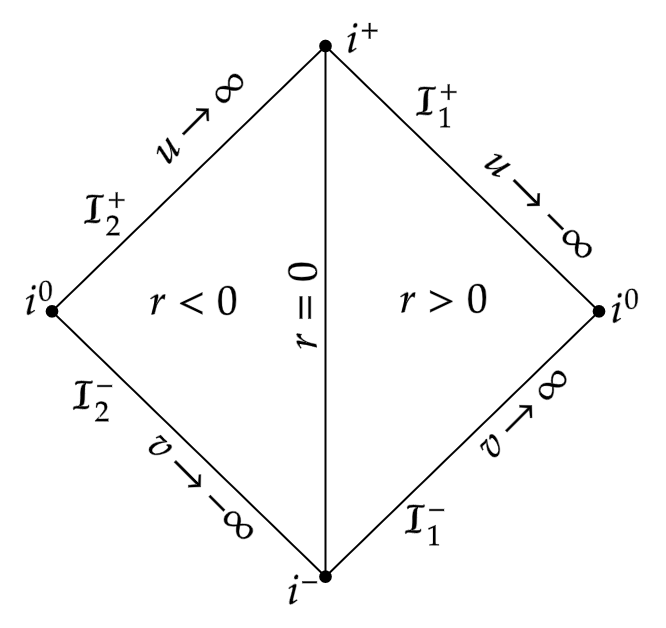}
    \caption{The Penrose diagram for the Ellis-Bronnikov wormhole spacetime. This compactified spacetime consists of two copies of compactified Minkowski spacetime, each corresponding to one of the two universes $r<0$ and $r>0$.}
    \label{fig:wormhole}
\end{figure}

In order to consider the boundary ersion of the Penrose property it is necessary to compactify this spacetime. We can do this using the same method as for Minkowski spacetime \cite[Section 3]{PCMD} to obtain the conformally related metric
\begin{equation}\label{eqn:wormholecompact}
    \overline{ds}^2=dT^2-d\chi^2-\left(1+\frac{a^2}{r^2}\right)\sin^2\chi d\omega^2_{d-1}
\end{equation}
Note that we have $r\in(-\infty,\infty)$, so the range of $\chi$ is $\chi\in[-\pi,\pi]$ (for Minkowski we had $r\in[0,\infty)$ and hence $\chi\in[0,\pi]$). We also have $T\in[-\pi,\pi]$ and $|T|\leq\pi-|\chi|$. This means we obtain a Penrose diagram which looks like two copies of the one obtained for Minkowski spacetime in \cite{PCMD}, one for each of the two ```universes'' $r<0$ and $r>0$. This is shown in Figure \ref{fig:wormhole}, where each point represents a manifold which is topologically a $(d-1)$-sphere. 

In this section (and in Section \ref{Hayward Metric}) we will drop the distinction between the two versions of the Penrose property since Theorem \ref{thm:equivalence} shows that they are equivalent. Of these two equivalent formulations, it will be more convenient to work with the non-timelike boundary version. 

The problem of timelike connecting points on $\mathcal{I}^-$ and $\mathcal{I}^+$ can be split into two cases. The first case consists of points which both lie in the same universe, so either $r=\infty$ at both or $r=-\infty$ at both. The second case consists of points in different universes, so one point is at $r=\infty$ and the other is at $r=-\infty$. We use subscripts 1 and 2 to denote the two universes $r>0$ and $r<0$ respectively.

We begin by dealing with the case where both points lie in the same universe, which we choose without loss of generality to be the one with $r>0$. For a curve restricted to this universe, we let $R$ denote its \textit{impact parameter} - the smallest value of $r$ attained along the curve. It will be helpful to make the following definition.

\begin{definition}\label{defn:timeofflight}
The \textit{time of flight} along a curve with endpoints is defined to be the difference between the retarded time at its future endpoint, denoted $u_\infty$, and the advanced time at its past endpoint, denoted $v_\infty$. 

Note that the time of flight along a curve need not be positive or finite (for example any timelike curve with future endpoint at $i^+$ would have infinite time of flight).
\end{definition}
For the Ellis-Bronnikov wormhole we use the retarded and advanced time co-ordinates $u=t-|r|$ and $v=t+|r|$ respectively.

\begin{prop}\label{prop:Mink}
Let $p\in\mathcal{I}_1^-$ and let $q\in\mathcal{I}_1^+$ be antipodal points such that any curve from $p$ to $q$ would have zero time of flight. Then 
\begin{equation}
    I^+(p)\cap\mathcal{I}_1^+=\mathcal{I}_1^+\setminus \left(J^-(q)\cap\mathcal{I}_1^+\right)
\end{equation}
That is, the only points in $\mathcal{I}_1^+$ which cannot be timelike connected to $p$ are the antipodal points where the time of flight would be $\leq0$.
\end{prop}

\textbf{Proof}: We begin by showing that, as their impact parameter $R\rightarrow\infty$, the future endpoints of null geodesics from $p$ tend to the antipodal point on the $(d-1)$-sphere and their time of flight tends to zero. 

Solving the geodesic equations, we find that the change in the retarded time co-ordinate $u=t-r$ along a null geodesic with impact parameter $R$ from the point where $r=R$ to its future endpoint at $r=\infty$ is
\begin{equation}
    u_\infty-u_R=\int_R^\infty\left[\left(1-\frac{R^2+a^2}{\rho^2+a^2}\right)^{-1/2}-1\right]d\rho
\end{equation}

Similarly, the change in the advanced time co-ordinate $v=t+r$ from the past endpoint to the point where $r=R$ is
\begin{equation}
    v_\infty-v_R=-\int_R^\infty\left[\left(1-\frac{R^2+a^2}{\rho^2+a^2}\right)^{-1/2}-1\right]d\rho
\end{equation}

Hence the time of flight along this null geodesic is
\begin{equation}
    \begin{split}
        u_\infty-v_\infty&=2\int_R^\infty\left[\left(1-\frac{R^2+a^2}{\rho^2+a^2}\right)^{-1/2}-1\right]d\rho+u_R-v_R\\
        &= 2R\int_1^\infty\left[\left(1-\frac{1+a^2/R^2}{x^2+a^2/R^2}\right)^{-1/2}-1\right]dx-2R\\
        &=\frac{2af(b)}{b}
    \end{split}
\end{equation}
where we have made the substitution $x=\rho/R$ and defined $b=a/R$ and $f(b)=\int_1^\infty\left[\left(\frac{x^2+b^2}{x^2-1}\right)^{1/2}-1\right]dx-1$. 

Now $f(0)=0$, so we can use l'H\^{o}pital's rule to determine the limit of $u_\infty-v_\infty$ as $b\rightarrow0$ (i.e. as $R\rightarrow\infty$). We have
\begin{equation}
    \begin{split}
        \lim_{R\rightarrow\infty}(u_\infty-v_\infty)&=\lim_{b\rightarrow0}2af'(b)\\
        &=\lim_{b\rightarrow0}2ab\int_1^\infty\left(x^2-1\right)^{-1/2}\left(x^2+b^2\right)^{-1/2}dx\\
        &=0
    \end{split}
\end{equation}
since 
\begin{equation}
    \int_1^\infty\left(x^2-1\right)^{-1/2}x^{-1}dx=\frac{\pi}{2}
\end{equation}

\begin{figure} 
    \centering
    \includegraphics[scale=0.3]{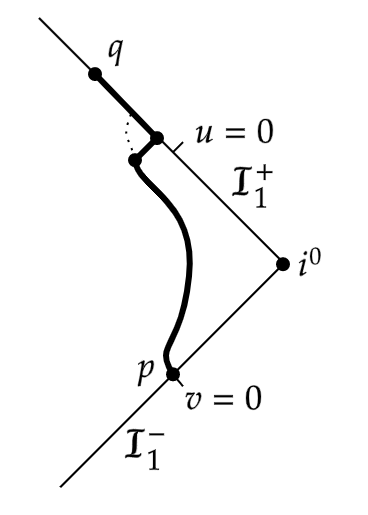}
    \caption{This figure shows the construction of a timelike path with strictly positive time of flight between $p\in\mathcal{I}^-_1$ and $q\in\mathcal{I}^+_1$. We follow a null geodesic in the same plane as $p$ and $q$ which reaches $\mathcal{I}^+_1$ at an earlier retarded time than $q$. Once we reach the same angular co-ordinate as $q$, we switch to an outgoing radial null geodesic until we reach $\mathcal{I}^+_1$ and then another radial null geodesic up $\mathcal{I}^+_1$ until we reach $q$. We then smooth this path (without changing its endpoints) so that it becomes timelike everywhere.}
    \label{fig:Penrosepath}
\end{figure}

We also need to check the angular change along a null geodesic with endpoints on $\mathcal{I}^-$ and $\mathcal{I}^+$. We choose co-ordinates so that our null geodesic lies in the equatorial plane. Then 
\begin{equation}
    \begin{split}
        \frac{d\phi}{dr}&=\frac{(R^2+a^2)^{1/2}}{(r^2-R^2)^{1/2}(r^2+a^2)^{1/2}}\\
        \implies\Delta\phi&=2(R^2+a^2)^{1/2}\int_R^\infty(r^2-R^2)^{-1/2}(r^2+a^2)^{-1/2}dr\\
        &\geq 2R\int_R^\infty r^{-1}(r^2-R^2)^{-1/2}dr\\
        &=\pi
    \end{split}
\end{equation}
We conclude that $p\in\mathcal{I}^-_1$ can be connected to any point $q\in\mathcal{I}^-_1$ where the time of flight is strictly positive using the following construction (illustrated in Figure \ref{fig:Penrosepath}).

Choose co-ordinates such that $p$ and $q$ lie in the equatorial plane. By the above result, we can choose a null geodesic from $p$ (lying in the equatorial plane) with sufficiently large impact parameter to ensure that it reaches $\mathcal{I}^+_1$ at an earlier retarded time than $q$. We follow this null geodesic until we have reached the $\phi$ co-ordinate of $q$ (reversing the angular momentum if necessary so that the required change in $|\phi|$ is at most $\pi$) before switching to an outgoing radial geodesic until we reach $\mathcal{I}_1^+$ and then another null geodesic up $\mathcal{I}_1^+$ until we reach $q$. This path can then be smoothed (without changing its endpoints) so that it becomes timelike everywhere \cite{DiffTop}. 

Next we consider points separated by a curve with non-positive time of flight. It is straightforward to see that antipodal points cannot be timelike connected by using a comparison argument similar to those developed in \cite{PCMD}. Indeed, we note that we can bound the compactified wormhole metric (equation (\ref{eqn:wormholemetric})) from above by the compactified Minkowski metric (which corresponds to setting $a=0$ in this metric).
\begin{equation}
    \overline{ds}^2_{E-B}\leq \overline{ds}^2_{Mink}
\end{equation}
This tells us that if two points on $\mathcal{I}_1^\pm$ cannot be connected using a Minkowski timelike curve then they cannot be connected using a curve which is timelike with respect to the wormhole metric either (since any such curve would necessarily be Minkowski timelike). From \cite[Proposition 3.3]{PCMD}, we conclude that in the wormhole spacetime, antipodal points separated by curves with time of flight $\leq0$ cannot be timelike connected. 

We now consider the case where $p$ and $q$ are non-antipodal. We once again choose co-ordinates so that these points lie in the equatorial plane. 

For $r\geq R$, we have
\begin{equation}\label{eqn:comparison}
    \begin{split}
    \overline{ds}^2&=dT^2-d\chi^2-\left(1+\frac{a^2}{r^2}\right)\sin^2\chi d\phi^2\\
    &\geq dT^2-d\chi^2-\left(1+\frac{a^2}{R^2}\right)\sin^2\chi d\phi^2\\
    &=dT^2-d\chi^2-\sin^2\chi d\overline{\phi}^2\\
    &=\overline{dg}_{Mink}^2
    \end{split}
\end{equation}
where $\overline{\phi}=\phi\left(1+\frac{a^2}{R^2}\right)^{1/2}\in\left[0,\pi\sqrt{1+\frac{a^2}{R^2}}\right)$. 

From \cite[Proof of Proposition 3.3]{PCMD}, we see that given any $R$ it is possible to connect any non-antipodal points (i.e. points in the equatorial plane with $\Delta\overline{\phi}<\pi$) using a Minkowski timelike curve which remains at $r>R$. Suppose the $\phi$ co-ordinate of such a curve is given by the function $\overline{\phi}(\tau)$, where $\tau$ is some affine parameter. The above inequality tells us that the same curve now with $\phi$ co-ordinate $\phi(\tau)=\overline{\phi}(\tau)\left(1+\frac{a^2}{R^2}\right)^{-1/2}$ will be timelike with respect to the compactified wormhole metric. This method allows us to timelike connect any points in the equatorial plane whose angular difference is
\begin{equation}
    \Delta\phi<\frac{\pi}{\sqrt{1+\frac{a^2}{R^2}}}
\end{equation}
By choosing $R$ sufficiently large, we conclude that all non-antipodal points can be timelike connected. \qed

\begin{figure} 
  \minipage{0.48\textwidth}
   \centering
        \includegraphics[height=6cm]{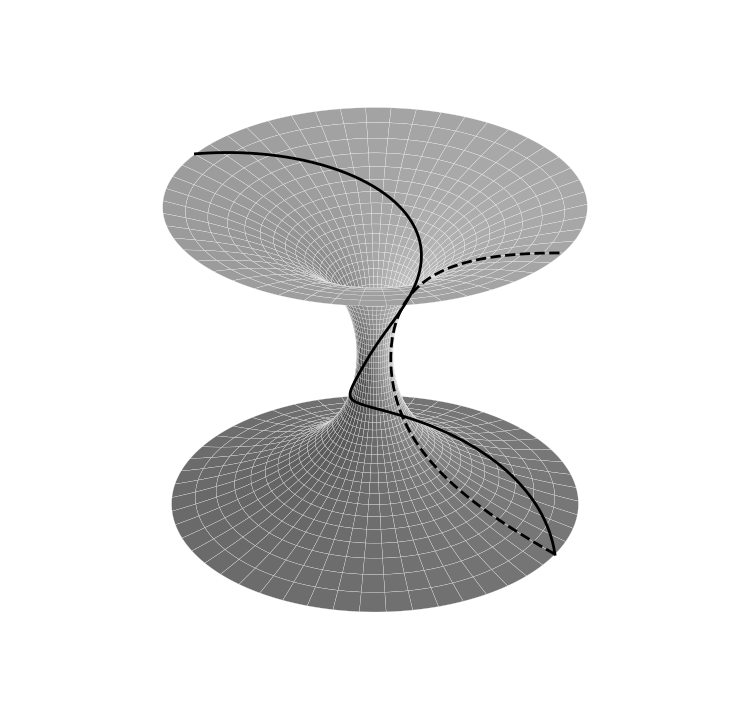}
    \caption{The dashed line shows a null geodesic with zero angular momentum ($h=0$) which passes through the throat. The case $0<h<a$ is shown as a solid line. This geodesic passes through the throat and intersects $\mathcal{I}_2^+$.
    \\
    \\}
    \label{fig:wormhole1}
    \endminipage
     \minipage{0.04\textwidth}
      \includegraphics[scale=0.05]{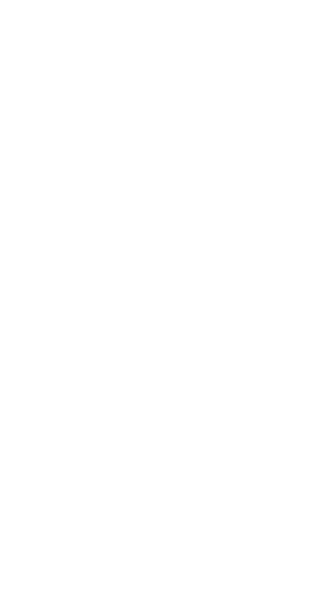}
      \endminipage
    \minipage{0.48\textwidth}
 \centering
    \includegraphics[height=6cm]{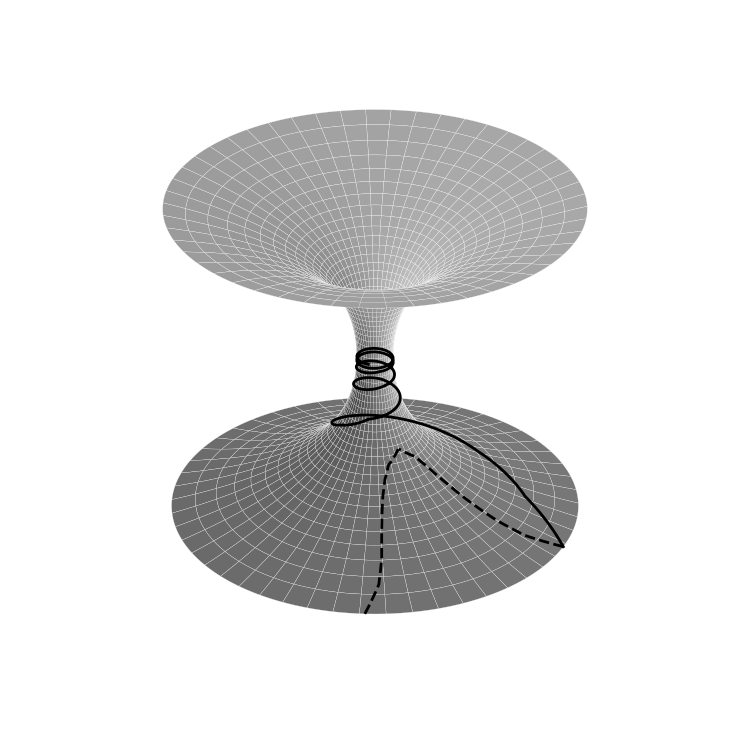}
    \caption{The solid line shows a null geodesic with angular momentum $h=a$. This geodesic asymptotes towards the throat at $r=0$ and has an endpoint at future timelike infinity $i^+$. The case $h>a$ is shown as a dashed line. This geodesic does not pass through the throat and has future endpoint on $\mathcal{I}_1^+$.}
    \medskip
    \label{fig:wormhole2}
    \endminipage
\end{figure}

Next we consider points on $\mathcal{I}^-$ and $\mathcal{I}^+$ which lie in different universes. As in previous constructions we will rely on null geodesics. We begin by asking which of these cross the wormhole throat and which remain in one universe.
\begin{lemma}\label{lemma:wormholecrossinglemma}
Let $\gamma$ be a null geodesic in the equatorial plane with past endpoint $p\in\mathcal{I}^-_1$. Then $\gamma$ crosses the wormhole throat of width $a$ if and only if it has angular momentum $h=(r^2+a^2)\dot{\phi}$ satisfying $h^2<a^2$ (where $\dot{}$ denotes differentiation with respect to $t$). If $h^2=a^2$ then the geodesic has an endpoint at future timelike infinity but does not cross the wormhole.
\end{lemma}
This is illustrated in figures \ref{fig:wormhole1} and \ref{fig:wormhole2} where paths are shown in the equatorial plane in a $t=$ constant slice of the spacetime.

\textbf{Proof}: The equation of motion in the radial direction (parametrised by $t$) is
\begin{equation}
\begin{split}
    \dot{r}^2&=1-\frac{h^2}{r^2+a^2}\\
    \implies r^2&\geq h^2-a^2
\end{split}
\end{equation}
So we see that if $h^2>a^2$, then the geodesic does not cross the wormhole throat at $r=0$ and is restricted to the $r>0$ universe. If $h^2=a^2$ then $\dot{r}\rightarrow0$ as $r\rightarrow0$. The retarded time co-ordinate diverges along this geodesic, so it has an endpoint at future timelike infinity $i^+$. If $h^2<a^2$, then we have $\dot{r}\neq0$ and the geodesic passes through $r=0$ and has future endpoint on $\mathcal{I}^+_2$. \qedsymbol

We are now in a position to consider exactly which points in $\mathcal{I}^-_1$ and $\mathcal{I}^+_2$ can be timelike connected.

\begin{prop}\label{prop:Penrosedifferentuniverses}
Let $p\in\mathcal{I}_1^-$ and $q\in\mathcal{I}_2^+$ and choose co-ordinates such that these points lie in the equatorial plane with $\phi=0$ at $p$ and $\phi=\phi_q$ at $q$. Let $u_{(h)}$ denote the value of the retarded time co-ordinate $u=t-|r|$ at the future endpoint of a null geodesic from $p$ in the equatorial plane with angular momentum $h=(r^2+a^2)\dot{\phi}$ (where $\dot{}$ denotes differentiation with respect to $t$). 
\begin{itemize}
    \item Suppose $q$ has retarded time less than or equal to $u_{(0)}$. Then $p$ cannot be timelike connected to $q$.
    \item Suppose $u>u_{(0)}$ at $q$. Then $p$ can be timelike connected to $q$ if and only if $\phi_q\in[-\frac{2h_q}{a}K\left(\frac{h_q}{a}\right),\frac{2h_q}{a}K\left(\frac{h_q}{a}\right)]$, where $K(k)$ denotes the complete elliptic integral of the first kind\footnote{The complete elliptic integral of the first kind is defined as $K(k)=\int_1^\infty\frac{dt}{\sqrt{(1-t^2)(1-k^2t^2)}}$ for $|k|<1$.} and $h_q$ denotes the angular momentum of a null geodesic from $p$ to $q$.
\end{itemize}
\end{prop}

\textbf{Proof}: By \cite[Lemma 3.1]{PCMD}, in order to timelike connect $p$ and $q$ it suffices to consider only curves which lie in the equatorial plane. 

\begin{figure} 
    \centering
    \includegraphics[scale=0.3]{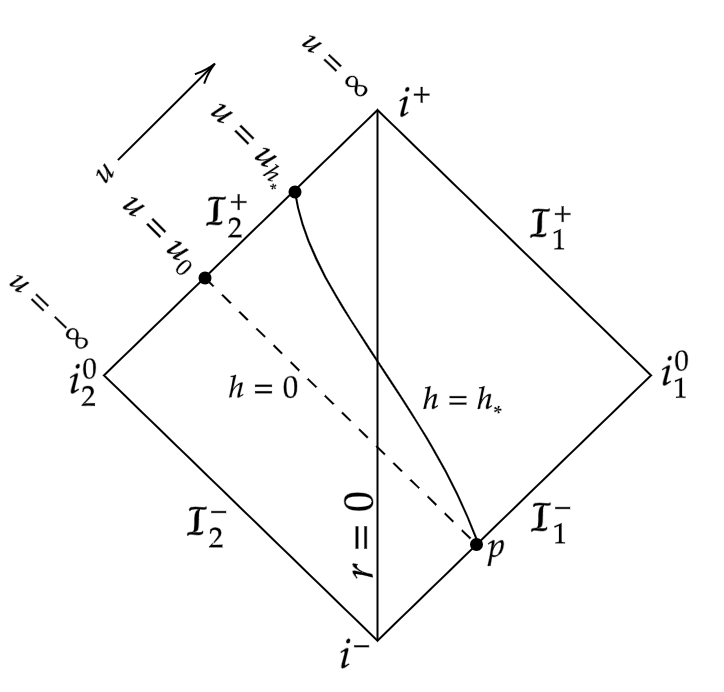}
    \caption{The dotted line shows a radial (i.e. $h=0$) null geodesic from $p$ passing through the wormhole throat and finishing at $u$ co-ordinate $u_{(0)}$. The point $p\in\mathcal{I}^-_1$ cannot be timelike connected to any points across the wormhole on $\mathcal{I}_2^+$ which have $u\leq u_{(0)}$. The solid line shows a null geodesic from $p$ with angular momentum $h_*$. It reaches $\mathcal{I}_2^+$ at $u$ co-ordinate $u_{(h_*)}$ and the angular change along it is exactly $\pi$. The point $p$ can be timelike connected to all points on $\mathcal{I}_2^+$ which have $u>u_{(h_*)}$. For points with $u_{(0)}< u\leq u_{(h_*)}$, the result depends on the angular separation from $p$ (see Proposition \ref{prop:Penrosedifferentuniverses}).}
    \label{fig:wormholepaths}
\end{figure}

By inspection of the Penrose diagram (Figure \ref{fig:wormholepaths}) and consideration of a radial null geodesic (i.e. a null geodesic with $h=0$), we can see immediately that no points on $\mathcal{I}_2^+$ with $u$ co-ordinate $u\leq u_{(0)}$ can be reached by a timelike curve from $p$. 

Now consider $q\in\mathcal{I}^+_2$ with $u$ co-ordinate $>u_{(0)}$ and assume without loss of generality that the $\phi$ co-ordinate of $q$ is $\phi_q\in[0,\pi]$ (otherwise reverse the sign of $h$ in what follows).

Once again consider a null geodesic from $p$ lying in the equatorial plane. Suppose this geodesic has angular momentum $h\in(0,a)$ and hence, by Lemma \ref{lemma:wormholecrossinglemma}, crosses the wormhole throat. The change in the $\phi$ co-ordinate along this geodesic, denoted $\Delta\phi$, is given by
\begin{equation}\label{eqn:deltaphiwormhole}
    \begin{split}
      \Delta\phi&=\int_{-\infty}^\infty\frac{hdr}{\left(r^2+a^2\right)^{1/2}\left(r^2+a^2-h^2\right)^{1/2}}\\
      &=\frac{2h}{a}K\left(\frac{h}{a}\right)
    \end{split}
\end{equation}

So $\Delta\phi$ is a strictly increasing continuous function of $\frac{h}{a}$ for $0\leq h<a$, with $\Delta\phi=0$ at $h=0$ and $\Delta\phi\rightarrow\infty$ as $h\rightarrow a$. 

The change in the retarded time co-ordinate $u$ along this geodesic, denoted $\Delta u$, is given by
\begin{equation}\label{eqn:deltauwormhole}
    \Delta u = 2\int_0^\infty\left(1-\sqrt{1-\frac{h^2}{r^2+a^2}}\right)dr
\end{equation}
So $\Delta u$ is an increasing continous function of $h$ for $0\leq h<a$, with $\Delta u=0$ at $h=0$ and $\Delta u\rightarrow\infty$ as $h\rightarrow a$. This tells us in particular that $u_{(0)}$ is equal to the $u$ co-ordinate of $p$.

Since $u>u_{(0)}$ at $q$, using equation (\ref{eqn:deltauwormhole}) we can use the inverse function theorem to write this $u$ co-ordinate as $u_{({h_q})}$ for some unique $h_q\in(0,a)$. If $0\leq\phi_q<\frac{2h_q}{a}K\left(\frac{h_q}{a}\right)$, then we can reach the point $q$ with a timelike curve from $p$ as follows:

\begin{itemize}
    \item Follow the null geodesic from $p$ in the equatorial plane with angular momentum $h_q$ until we reach $\phi=\phi_q$ (which we do before reaching $\mathcal{I}_2^+$ since $\Delta\phi=\frac{2h_q}{a}K\left(\frac{h_q}{a}\right)$ along the full null geodesic).
    \item Complete the path to $q\in\mathcal{I}^+_2$ with the timelike curve which follows the same path as the null geodesic in the first step except with $\dot{\phi}=0$.
\end{itemize}

This curve can then be smoothed to form a path from $p$ to $q$ which is timelike everywhere \cite{DiffTop}. In particular, all points with $u$ co-ordinate $u_{({h_q})}> u_{(h_*)}$ can be reached by a timelike curve from $p$, where $h_*$ is the unique value in $(0,a)$ satisfying $\frac{2h_*}{a}K\left(\frac{h_*}{a}\right)=\pi$.

Now suppose $q$ has $u$ co-ordinate $u_{({h_q})}\in(u_{(0)},u_{(h_*)}]$ and $\phi$ co-ordinate $\phi_q\geq\frac{2h_q}{a}K\left(\frac{h_q}{a}\right)$. Suppose that a timelike curve from $p$ to $q$ exists. Since the causal diamond $J^+(p)\cap J^-(q)$ (considered as a subset of the compactified spacetime) is non-empty, compact and strongly causal, there exists a causal geodesic from $p$ to $q$ \cite[theorem 1]{Seifert}. If this geodesic were null then the change in $\phi$ along it would be $\frac{2h_q}{a}K\left(\frac{h_q}{a}\right)$ (up to sign). The change in $\phi$ along a timelike geodesic with the same endpoints would therefore be strictly less than $\frac{2h_q}{a}K\left(\frac{h_q}{a}\right)$. We conclude that there is no timelike curve between $p$ and $q$. \qedsymbol

\subsection{Hayward Metric}\label{Hayward Metric}
The Hayward spacetime describes a non-singular black hole. The metric in $3+1$ dimensions is 
\begin{equation}
\begin{split}
    ds^2&=V(r)dt^2-\frac{dr^2}{V(r)}-r^2-d\omega_2^2\\
    V(r)&=1-\frac{2mr^2}{r^3+2l^2m}
    \end{split}
\end{equation}
where $m$ is the ADM mass and $l$ is a length-scale parameter. The metric has the same asymptotic form as the Schwarzschild metric but there is no curvature singularity at $r=0$. The construction used by Penrose in \cite{Penrose} (see also the proof of theorem 4.2 in \cite{PCMD}) shows that the Penrose property is a property of spacetimes near $i^0$ (i.e. at large $r$). We therefore expect the Penrose property to be satisfied when $m>0$ but not when $m\leq0$ (as was the case for the Schwarzschild metric). 
\begin{prop}
The Penrose property is satisfied in the $3+1$ dimensional Hayward spacetime when $m>0$ but not when $m\leq0$
\end{prop}
\textbf{Proof}: First note that the $m=0$ case is Minkowski spacetime. This was shown in \cite[Section IV]{Penrose} (also \cite[Theorem 2.6]{PCMD}) not to satisfy the Penrose property.

For the $m\neq0$ cases, it will be necessary to compactify the Hayward spacetime. We define the tortoise co-ordinate by
\begin{equation}
\begin{split}
    dr_*&=\frac{dr}{V(r)}\\
    \implies r_*&=r+2m\log(r/2m-1)+O(r^{-3})
\end{split}
\end{equation}
Crucially, this co-ordinate contains the same logarithmic term as Schwarzschild in $3+1$ dimensions. The higher order terms decay near $i^0$ and will not be important. In fact, it is straightforward to check that the proofs given in \cite[Sections 4.21 and 4.22]{PCMD} still hold and hence that the Penrose property is satisfied when $m>0$ but not when $m\leq0$. \qedsymbol

We comment at this point that if we allow our curves to cross event horizons then the two versions of the Penrose property may no longer be equivalent. In particular, the Hayward metric is an example of a spacetime which contains a horizon but no singularity. If this spacetime is extended across $r=0$ then it would seem plausible that we could find two endless timelike curves, one of which crosses the event horizon and passes into this extension, which cannot be timelike connected. 

\section{Conclusion}
In order to rule out a formulation of quantum gravity based on a fixed background Minkowski spacetime, Penrose showed in \cite{Penrose} that it was sufficient to find one physically relevant spacetime which satisfies the non-timelike boundary version of the Penrose property. For asymptotically de Sitter spacetimes, the Schwarzschild-de Sitter black hole satisfies this property and hence rules out a ``$SO(d+1,1)$ covariant'' construction of quantum gravity based on a background de Sitter spacetime. Note that unlike in the asymptotically flat case, such a construction is ruled out in $d+1$ dimensions for any $d\geq3$. For asymptotically anti-de Sitter spacetimes we can rule out a quantum gravity construction if we are able to find a physically relevant spacetime satisfying the timelike boundary version of the Penrose property (Definition \ref{defn:PenrosePropertyAdS}). However, the spacetime we have found which satisfies this is the negative mass Schwarzschild-AdS spacetime which we may regard as not being physically relevant. Furthermore, the Theorem of Gao and Wald (Theorem \ref{thm:GaoWald}) shows that this property fails for spacetimes which focus null geodesics. These are spacetimes we would like to regard as physical since they are associated with spacetimes containing positive energy densities. As a result, we are unable to rule out a quantum gravity construction based on a fixed background anti-de Sitter spacetime.

We have also considered the Penrose property in product spacetimes which are composed of a Lorentzian manifold and a compact Riemannian manifold. We found that both versions of the Penrose property were satsified in this product spacetime if and only if they are satisfied in the Lorentzian spacetime alone. This was perhaps surprising since considering a product spacetime may appear to give a more restrictive condition. We are therefore able to rule out a quantum gravity construction based on a fixed background spacetime consisting of the product of Minkowski spacetime with any compact Riemannian manifold.

Finally, we considered the Penrose property in two
examples of asymptotically flat spacetimes highlighting properties which were not encountered in \cite{PCMD}. We showed how similar techniques to those employed in \cite{PCMD} allowed us to treat these examples successfully.  The wormhole example showed that the Penrose property can be more complicated if we consider spacetimes with more than one asymptotically flat end since (as for asymptotically de Sitter spacetimes) we cannot simply rely on the asymptotic behaviour of the metric. The example of the Hayward metric, a non-flat spacetime with no singularities, was straightforward and highlighted the fact that the Penrose property is a property of neighbourhoods of spatial infinity and in particular does not require the existence of curvature singularities.
\subsection*{Acknowledgements} I am grateful to Maciej Dunajski, Roger Penrose, Harvey Reall, Paul Tod, and Claude Warnick for useful discussions and to Michael Cameron for his help with the figures. I would also like to thank the anonymous referees for their extremely helpful comments. I am supported St. John's College, Cambridge for their support through a College Scholarship from the Todd/Goddard Fund.

\end{document}